\documentclass{sig-alternate-05-2015}

\usepackage{times}
\usepackage{amsmath,amssymb}
\usepackage{latexsym}
\usepackage{graphicx}
\usepackage{comment}
\usepackage{caption}
\usepackage{subcaption}
\usepackage{tabularx}
\usepackage[pdfstartview=FitH]{hyperref} %generate bookmark and reference hyperlink, fit pdf content to the window
\usepackage[open,openlevel=2,numbered]{bookmark} 

\usepackage{color}

\newcommand{\floor}[1]{\lfloor #1 \rfloor}

\begin{document}

% Copyright
% \setcopyright{acmcopyright}
%\setcopyright{acmlicensed}
%\setcopyright{rightsretained}
%\setcopyright{usgov}
%\setcopyright{usgovmixed}
%\setcopyright{cagov}
%\setcopyright{cagovmixed}

% DOI
\doi{10.475/123_4}

% ISBN
\isbn{123-4567-24-567/08/06}

%Conference
\conferenceinfo{Under review}{}

\acmPrice{\$15.00}

%
% --- Author Metadata here ---
% \conferenceinfo{WOODSTOCK}{'97 El Paso, Texas USA}
%\CopyrightYear{2007} % Allows default copyright year (20XX) to be over-ridden - IF NEED BE.
%\crdata{0-12345-67-8/90/01}  % Allows default copyright data (0-89791-88-6/97/05) to be over-ridden - IF NEED BE.
% --- End of Author Metadata ---

\title{DeepCas: an End-to-end Predictor of Information Cascades}

\numberofauthors{1} %  in this sample file, there are a *total*
\author{
% 1st. author
 \alignauthor
 Cheng Li$^{1}$, Jiaqi Ma$^{1}$, Xiaoxiao Guo$^{2}$, Qiaozhu Mei$^{1,2}$\\
 		\affaddr{$^{1}$School of Information, University of Michigan, Ann Arbor, MI, USA}\\
 		\affaddr{$^{2}$Department of EECS, University of Michigan, Ann Arbor, MI, USA}\\
        \affaddr{\{lichengz, \ jiaqima, \ guoxiao, \ qmei\}@umich.edu}
}
% There's nothing stopping you putting the seventh, eighth, etc.
% author on the opening page (as the 'third row') but we ask,
% for aesthetic reasons that you place these 'additional authors'
% in the \additional authors block, viz.

% Just remember to make sure that the TOTAL number of authors
% is the number that will appear on the first page PLUS the
% number that will appear in the \additionalauthors section.

\maketitle
\begin{abstract}
Information cascades, effectively facilitated by most social network platforms, are recognized as a major factor in almost every social success and disaster in these networks. Can cascades be predicted? While many believe that they are inherently unpredictable, recent work has shown that some key properties of information cascades, such as size, growth, and shape, can be predicted by a machine learning algorithm that combines many features. These predictors all depend on a bag of hand-crafting features to represent the cascade network and the global network structure. Such features, always carefully and sometimes mysteriously designed, are not easy to extend or to generalize to a different platform or domain. 

Inspired by the recent successes of deep learning in multiple data mining tasks, we investigate whether an end-to-end deep learning approach could effectively predict the future size of cascades. Such a method automatically learns the representation of individual cascade graphs in the context of the global network structure, without hand-crafted features and heuristics. 
We find that node embeddings fall short of predictive power, and it is critical to learn the representation of a cascade graph as a whole. We present algorithms that learn the representation of cascade graphs in an end-to-end manner, which significantly improve the performance of cascade prediction over strong baselines that include feature based methods, node embedding methods, and graph kernel methods. Our results also provide interesting implications for cascade prediction in general. 

\end{abstract}

%%
%% The code below should be generated by the tool at
%% http://dl.acm.org/ccs.cfm
%\begin{CCSXML}
%<ccs2012>
%  <concept_id>10010520.10010553.10010562</concept_id>
%  <concept_desc>Computer systems organization~Embedded systems</concept_desc>
%  <concept_significance>500</concept_significance>
%  <concept_id>10010520.10010575.10010755</concept_id>
%  <concept_desc>Computer systems organization~Redundancy</concept_desc>
%  <concept_significance>300</concept_significance>
%  <concept_id>10010520.10010553.10010554</concept_id>
%  <concept_desc>Computer systems organization~Robotics</concept_desc>
%  <concept_significance>100</concept_significance>
%  <concept_id>10003033.10003083.10003095</concept_id>
%  <concept_desc>Networks~Network reliability</concept_desc>
%  <concept_significance>100</concept_significance>
%</ccs2012>  
%\end{CCSXML}
%
%\ccsdesc[500]{Computer systems organization~Embedded systems}
%% End generated code
%%
%
%%
%%  Use this command to print the description
%%
%\printccsdesc
%
%% We no longer use \terms command
%%\terms{Theory}
%
%\keywords{ACM proceedings; \LaTeX; text tagging}

\section{Introduction}
\label{sec:intro}
% TODO: motivation for prediction of diffusion: marketing
%With the rapid development of information technology, we accumulated huge amount of graph-structured data which forms \textit{information networks}, e.g., World Wide Web, online social networks, academic paper networks, etc. Information networks can be characterized by the interplay between heterogeneous content with a complex underlying network structure. Information flows and influences each other involved in the networks, which is sometimes desired -- like the spread of innovation in academic networks or viral marketing -- while sometimes undesired -- like panic or rumors.\cite{easley2010networks} An important research topic related to information diffusion, which is called \textit{cascade prediction}, is to predict, at the early stage of the cascade, whether a piece of information will become popular in future. 

Most modern social network platforms are designed to facilitate fast diffusion of information. Information cascades are identified to be a major factor in almost every plausible or disastrous social network phenomenon, ranging from viral marketing, diffusion of innovation, crowdsourcing, rumor spread, cyber violence, and various types of persuasion campaigns. 

If cascades can be predicted, one can make wiser decisions in all these scenarios. For example, understanding which types of Tweets will go viral helps marketing specialists to design their strategies; %estimating the growing popularity of research topics helps scientists identify promising directions; 
predicting the potential influence of a rumor enables administrators to make early interventions to avoid serious consequences. A prediction of cascade size benefits business owners, investors, journalists, policy makers, national security, and many others. 

Can cascades be predicted? While many believe that cascades are inherently unpredictable, recent work has shown that some key properties of information cascades, such as size, growth, and shape, can be predicted through a mixture of signals \cite{cheng2014can}. Indeed, 
%the prediction of certain types of cascades has been studied in various online social networks. C
%<<<<<<< HEAD
%cascades of microblogs/Tweets~\cite{yu2015micro,weng2014predicting,zhao2015seismic,jenders2013analyzing,cui2013cascading,guille2012predictive}, photos~\cite{cheng2014can}, videos~\cite{bauckhage2015viral} and academic papers ~\cite{shen2014modeling} are proved to be predictable to some extent. In most of these studies, cascade prediction is cast as classification or regression problems and be solved with machine learning techniques that incorporate many features. On one hand, many of these features are specific to the particular platform or the particular type of information being diffused. For example, whether a photo was posted with a caption is shown to be predictive of how widely it spread on Facebook \cite{cheng2014can}; specific wording on Tweets is shown to help them gain more retweets \cite{tan2014effect}. These features are indicative but cannot be generalized to other platforms or to other types of cascades. On the other hand, a common set of signals, those extracted from the network structure of the cascade or the broader social network, are reported to be predictive by multiple studies \cite{cheng2014can,yu2015micro,weng2014predicting}. 
%
%=======
cascades of microblogs/Tweets~\cite{yu2015micro,weng2014predicting,zhao2015seismic,jenders2013analyzing,cui2013cascading,guille2012predictive}, photos~\cite{cheng2014can}, videos~\cite{bauckhage2015viral} and academic papers ~\cite{shen2014modeling} are proved to be predictable to some extent. In most of these studies, cascade prediction is cast as classification or regression problems and be solved with machine learning techniques that incorporate many features \cite{weng2014predicting,cheng2014can,cui2013cascading,jenders2013analyzing}. On one hand, many of these features are specific to the particular platform or the particular type of information being diffused. For example, whether a photo was posted with a caption is shown to be predictive of how widely it spread on Facebook \cite{cheng2014can}; specific wording on Tweets is shown to help them gain more retweets \cite{tan2014effect}. These features are indicative but cannot be generalized to other platforms or to other types of cascades. On the other hand, a common set of features, those extracted from the network structure of the cascade, are reported to be predictive by multiple studies~\cite{cheng2014can,yu2015micro,weng2014predicting}. 
%>>>>>>> 31e94097ce37e1ff98bc4d3571748dbccaa8429c

% TODO: comparison to existing work, from deep perspective. why deep? hand-crafted feature may not capture some information, interaction between g and G. end-to-end

%Cascade prediction can be naturally modeled as classification or regression problems and be solved with machine learning techniques. Most existing methods tackling this problem can be split into two categories. 

%One way to predict cascades through the network structure is to design generative models of the cascading process, most of which are some kinds of stochastic processes in the network (such as contagions or random walks). Parameters of these models are estimated by fitting the observed data, and the model generates a prediction of a new cascade through simulation or inference  \cite{bauckhage2013mathematical,yu2015micro,zhao2015seismic,shen2014modeling}. These models %do give good explanations of the mechanism behind information cascades, which however is 
%are usually an oversimplified version of the reality. As a result, these diffusion models are generally underperforming in real prediction tasks. 

%One category first 
%A common practice in literature is to transform the network structures into hand-crafted features and then trains standard discriminative predictors, e.g. linear regression or decision trees, using historical observations as supervision \cite{weng2014predicting,cheng2014can,cui2013cascading,jenders2013analyzing}. 
Many of these features are carefully designed based on the prior knowledge from network theory and empirical analyses, such as centrality of nodes, community structures, tie strength, and structural holes. There are also ad hoc features that appear very predictive, but their success is intriguing and sometimes magical. For example, Cheng et al. \cite{cheng2014can} found that one of the most indicative feature to the growth of a cascade is whether any of the first a few reshares are not directly connected to the root of the diffusion. 

We consider this as a major deficiency of these machine learning approaches: their performance heavily depends on the feature representations, yet there is no common principle of how to design and measure the features. Is degree the correct measure of centrality? Which algorithm should we use to extract communities, out of the hundreds available? How accurately can we detect and measure structural holes? How do we systematically design those ``magical'' features, and how do we know we are not missing anything important? Chances are whichever decisions we make we'll be losing information and making mistakes, and these mistakes will be accumulated and carried through to the prediction. 

%Another category builds generative models (mostly are some kinds of random processes) about diffusion process and then fit the observed data into the models to predict the later cascades\cite{bauckhage2013mathematical,yu2015micro,zhao2015seismic,shen2014modeling}. The first category suffers the common problem of standard machine learning approaches: the prediction performance heavily depends on the feature representations. What makes things worse in this problem is that different kinds of networks have heterogeneous data and not some features of one network may not be available in another network. There so far has not been a common principle for the feature design in cascade prediction problem. The generative models in the second category suffer from oversimplifying the mechanism of information diffusion. Researchers have been developing more and more complex diffusion models but the real diffusion process is so complex %TODO: add a citation prove the complexity 
%that the hand-crafted models still cannot capture it. 

%The result of Bauckhage et al.\cite{bauckhage2013mathematical}, though they claim that there is a general mechanism govern the cascade process, shows that different kinds of distributions fit different kinds of cascade best, which implies that precisely modeling the cascade process with hand-crafted distributions can be very hard. 

Can one overcome this deficiency? The recent success of \textit{deep learning} in different fields inspires us to investigate an end-to-end learning system for cascade prediction, a system that pipes all the way through the network structures to the final predictions without making arbitrary decisions about feature design. %In general, deep learning tackles these problems by automatically learning both the feature representation and the diffusion model from labeled data. There have been huge successes of deep learning applications in the areas of computer vision, speech recognition and natural language processing. 
Such a deep learning pipeline is expected to automatically learn the representations of the input data (cascade graphs in our case) that are the most predictive of the output (cascade growth), from a finer-granularity to increasingly more abstract representations, and allow the lower-level representations to update based on the feedback from the higher levels. A deep neural network is particularly good at learning a nonlinear function that maps these representations to the prediction, in our case the future size of a cascade. While deep learning models have shown their great power of dealing with image, text, and speech data, how to design a suitable architecture to learn the representations of \textit{graphs} remains a major challenge. %DeepWalk~\cite{perozzi2014deepwalk} tries to learn the representation of nodes in a graph by deploying a truncated random walk on social networks and treats the sampled paths as natural language sentences and the nodes as words. % LINE\cite{tang2015line} further interprets that the representation learned by DeepWalk actually preserves the proximity between nodes in terms of their shared neighbors. This is similar to one of the core ideas of deep learning models for natural language processing, the distributional representation, which means representing a word by means of its context. 
%TODO: more related work?
In the context of cascade prediction, the particular barrier is how to go from representations of nodes to representing a cascade graph as a whole. %Simply averaging node embeddings learned through DeepWalk \cite{perozzi2014deepwalk} or LINE \cite{tang2015line} results in a significant loss of structural information; graph kernels \cite{shervashidze2009efficient,shervashidze2011weisfeiler,borgwardt2005shortest,gartner2003graph,yanardag2015deep} could represent the entire graph but are hard to identify individual nodes, which is critical in cascade prediction. Indeed, knowing who retweets a message helps us guess how far it reaches.  %Simply taking the average of node embeddings learned from DeepWalk as the graph representation could result in a significant loss of structural information of the graphs. On the other hand, while graph kernels methods~\cite{xxx} are good at modeling the graph structure, it is hard for them to incorporate node identity information, which is crucial for diffusion prediction. A tweet is more likely to go viral if it is retweeted by President Obama than an average person.

We present a novel, end-to-end deep learning architecture named the \textit{DeepCas}, which first represents a cascade graph as a set of cascade paths that are sampled through multiple random walks processes. Such a representation  not only preserves node identities but also bounds the loss of structural information. Analogically, cascade graphs are represented as documents, with nodes as words and paths as sentences. The challenge is how to sample the paths from a graph to assemble the ``document,'' which is also automatic learned through the end-to-end model to optimize the prediction of cascade growth. Once we have such a ``document'' assembled, deep learning techniques for text data could be applied in a similar way here. We evaluate the performance of the proposed method using real world information cascades in two different domains, Tweets and scientific papers. DeepCas is compared with multiple strong baselines, including feature based methods, node embedding methods, and graph kernel methods. DeepCas significantly improves the prediction accuracy over these baselines, which provides interesting implications to the understanding of information cascades. 

%, while nodes as words. However 
%Unlike text documents where text are already written by humans, how to sample paths from a graph remains a challenge. We present a method to represent the diffusion graph using random walk, and manage to learn how to walk in an end-to-end manner, so that it could best predict the future diffusion size. Nodes are embedded into a hidden space, so that the embedding of the nodes could carry neighboring information globally.

\section{Related work}
\label{sec:related}
In a networked environment, people tend to be influenced by their neighbors' behavior and decisions~\cite{easley2010networks}. Opinions, product advertisements, or political propaganda could spread over the network through a chain reaction of such influence, a process known as the \textit{information cascade}%, is defined and developed for such processes
~\cite{welch1992sequential,bikhchandani1992theory,banerjee1992simple}. %Identifying and understanding which kind of information, at the early-stage of the cascade, will go viral is important in multiple domains, e.g. rumor detection, marketing, and diffusion of innovation. 
%% In literature, various social network properties have been connected to the outcome of information cascades, such as centrality, density, tie strength, communities, and structural holes~\cite{easley2010networks,yang2015rain}. 
We present the first deep learning method to predict the future size of information cascades.

\subsection{Cascade Prediction} 
%Cascade prediction in online social networks has been studied in various of scenarios and modeled in different ways. 
Cascades of particular types of information are empirically proved to be predictable to some extent, including Tweets/microblogs~\cite{yu2015micro, weng2014predicting,jenders2013analyzing,cui2013cascading,guille2012predictive, zhao2015seismic}, photos~\cite{cheng2014can}, videos~\cite{bauckhage2015viral} and academic papers~\cite{shen2014modeling}. In literature, cascade prediction is mainly formulated in two ways. One treats cascade prediction as a classification problem~\cite{weng2014predicting,jenders2013analyzing,cheng2014can,cui2013cascading}, which predicts whether or not a piece of information will become popular and wide-spread (above a certain threshold). The other formulates cascade prediction as a regression problem, which predicts the numerical properties (e.g., size) of a cascade in the future~\cite{weng2014predicting,tsur2012s}. This line of work can be further categorized by whether it outputs the final size of a cascade~\cite{zhao2015seismic} or the size  as a function of time (i.e., the growth of the cascade) ~\cite{yu2015micro}. Either way, most of the methods identified temporal properties, topological structure of the cascade at the early stage, root and early adopters of the information, and the content being spread as the most predictive factors. 

These factors are utilized for cascade prediction in two fashions. The first mainly designs generative models of the cascade process based on temporal or structural features, which can be as simple as certain macroscopic distributions (e.g., of cascade size over time) \cite{lerman2010information, bauckhage2013mathematical}, or stochastic processes that explain the microscopic actions of passing along the information \cite{yu2015micro}. These generative models make various strong assumptions and oversimplify the reality. As a result, they generally underperform in real prediction tasks. 
%% Bauckhage et al.\cite{bauckhage2013mathematical} assumed that there is a general mechanism governing the cascade process, but their results showed that different types of distributions fit different types of cascades, which implies that precisely modeling the cascade process is very hard. 

%There are two fashions in the methods of cascade prediction. 

Alternatively, these factors may be represented through handcrafted features, which are extracted from the data, combined, and weighted by discriminative machine learning algorithms to perform the classification or the regression tasks~\cite{weng2014predicting,cheng2014can,jenders2013analyzing,cui2013cascading}. %The second fashion is to build generative parametric models illustrating the cascade process and then fit the model with the data\cite{bauckhage2013mathematical,yu2015micro,zhao2015seismic,shen2014modeling}. 
Most work in this fashion uses standard supervised learning models (e.g. logistic regression, SVM, or random forests), the performance of which heavily rely on the quality of the features. In general, there is not a principled and systematic way to design these features. Some of the most predictive features are tied to particular platforms or particular cascades and are hard to be generalized, such as the ones mentioned in the Section~\ref{sec:intro}. Some features are closely related to the structural properties of the social network, such as degree\cite{cheng2014can,weng2014predicting}, density\cite{cheng2014can,guille2012predictive}, and community structures \cite{weng2014predicting}. These features could generalize over domains and platforms, but many may still involve arbitrary and hard decisions in computation, such as what to choose from hundreds of community detection algorithms available \cite{fortunato2010community} and how to detect structural holes \cite{yang2015rain}. Besides, there are also heuristic features that perform very well in particular scenarios but it is hard to explain why they are designed as is.

Our work differs from this literature as we take an end-to-end view of cascade prediction and directly learn the representations of a cascade without arbitary feature design. 
%We are mostly interested in predicting the future size of cascades at their early stage, as predictions by then are the most valuable and challenging. At later stages the prediction becomes easier (because observations are richer and temporal features become dominating factors). 
We focus on the structures (including node identities) of cascades as temporal and content information is not always available. In fact, content features are reported to be much weaker predictors than structural features \cite{cheng2014can}. Using temporal signals to predict future trend is a standard problem in time series, which is less interesting in this scope. 

%Previous work\cite{yu2015micro,zhao2015seismic,shen2014modeling} in the second fashion mainly model the cascade process as some kinds of distributions. %The result of Bauckhage et al.\cite{bauckhage2013mathematical}, though they claim that there is a general mechanism govern the cascade process, shows that different kinds of distributions fit different kinds of cascade best, which implies that precisely modeling the cascade process with handcraft distributions can be very hard. 

\begin{figure*}[ht]
\centering
\includegraphics[width=0.95\textwidth]{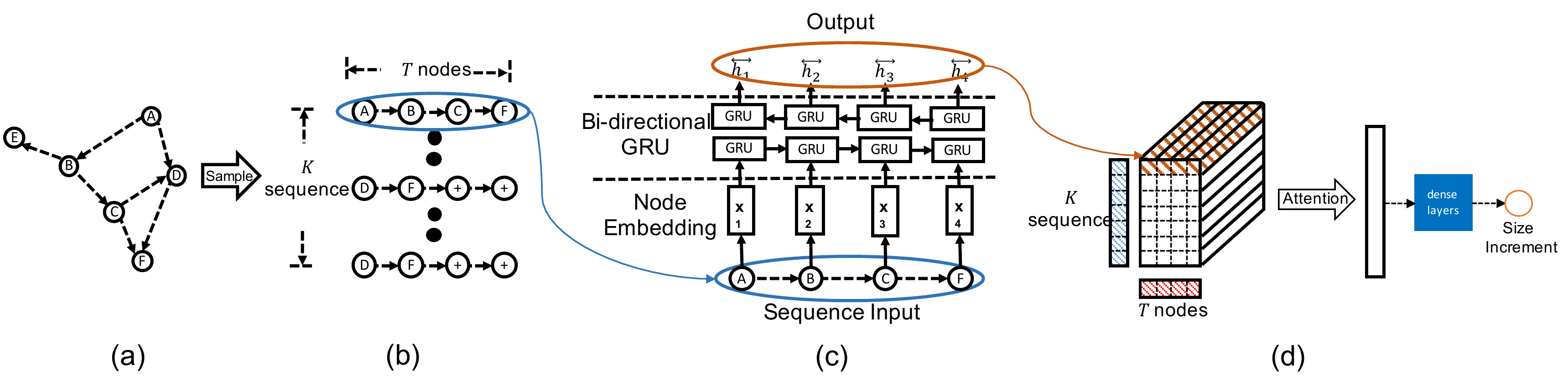}
\caption{The end-to-end pipeline of DeepCas.}
\label{fig:framework}
\end{figure*}

\subsection{Learning the Representation of Graphs}

Our work is also related to the literature of representation learning for graphs. Networks are traditionally represented as affiliation matrices or discrete sets of nodes and edges. Modern representation learning methods attempt to represent nodes as high-dimensional vectors in a continuous space (a.k.a., node embeddings) so that nodes with similar embedding vectors share similar structural properties (e.g., \cite{perozzi2014deepwalk, tang2015line, grovernode2vec}). Rather than learning the representation of each node, recent work also attempts to learn the representation of subgraph structures~\cite{niepert2016learning,narayanan2016subgraph2vec,yanardag2015deep,yanardag2015structural}. Much of this work is inspired by the huge success of representation learning and deep learning applied to various domains such as text \cite{bengio2003neural} and image \cite{krizhevsky2012imagenet}. For example, DeepWalk \cite{perozzi2014deepwalk} makes an analogy between the nodes in networks and the words in natural language and uses fixed-length random walk paths to stimulate the ``context'' of a node so that node representations can be learned using the same method of learning word representations~\cite{mikolov2013efficient}. The representation of a graph can then be calculated by averaging the embeddings of all nodes. 

Another line of related work comes from the domain of graph kernels, which computes pairwise similarities between graphs ~\cite{borgwardt2005shortest,gartner2003graph,shervashidze2011weisfeiler}. For example, the Weisfeiler-Lehman subtree kernel (\textbf{WL}) \cite{shervashidze2011weisfeiler} computes the graph similarity based on the sub-trees in each graph. Some studies have applied deep learning techniques to improve graph kernels~\cite{yanardag2015deep,subgraph2vec}. Though graph kernels are good at extracting structural information from a graph, it is hard for them to incorporate node identity information. %Relying on node identities to compute graph similarities could cause serious sparsity problem.

Another analogy connects graph structures to images. Motivated by representation learning of images, the topological structures of networks are first represented using locally connected regions~\cite{niepert2016learning}, spectral methods~\cite{defferrard2016convolutional}, and heat kernel signatures \cite{li2016deepgraph}, which could be passed through convolutional neural networks. These approaches are insensitive to orders of nodes and have an advantage of generating the same representation for isomorphic graphs. As the expense, it is also hard to incorporate the identities of nodes. %, which are critical to the prediction of cascades. 

Starting in next section, we present a novel end-to-end architecture that learns the representation of cascade graphs to optimize the prediction accuracy of their future sizes. %We start with a similar strategy as DeepWalk to sample a list of paths from a cascade graph but automatically learns the lengths and counts of the paths. Instead of averaging the node embeddings, a principled gated neural network is used to encode a  cascade graph as a whole. The entire learning and prediction procedure is handled in an end-to-end manner without arbitrary feature design. 

%[TODO: add related work about node embeddings]

%Recently, with the huge success of deep learning applications in various of areas, researchers start to develop deep representation of networks. 
%One branch of such work is to represent nodes with vectors by means of their neighbors, where an analogy between the nodes in networks and the words in natural language is made by treating the sample of paths in networks as the sentences. Another branch, rather than learning the representation of each node, learn the representation of subgraph structures\cite{niepert2016learning,narayanan2016subgraph2vec,yanardag2015deep,yanardag2015structural}. Deep learning methods have shown very promising power of approximating complex functions from enough amount of data. As either handcraft features or handcraft diffusion process functions are not robust, we are motivated to build a model learning the representation of networks and predictive functions simultaneously from the cascade data. 

\renewcommand{\vec}[1]{\mathbf{#1}}

\section{Method}

In reality, we observe snapshots of the social network but may or may not observe the exact time when nodes and edges are introduced. Similarly, we may observe snapshots of a cascade but not its complete history. In other words, at a given time we know who have adopted the information but not when or through whom the information was passed through ~\cite{cheng2014can} (e.g., we know who cited a paper but not when and where she found the paper). Below we define the problem so that it is closely tied to the reality.

\label{sec:method}

\subsection{Problem Definition}
\label{sec:problem}
Given a snapshot of a social network at time $t_0$, denote it as $\mathcal{G}=(V,E)$ where $V$ is the set of nodes and $E\subset V\times V$ is the set of edges. A node $i\in V$ represents an actor (e.g., a user in Twitter or an author in the academic paper network) and an edge $(i,j)\in E$ represents a relationship tie (e.g., retweeting or citation) between node $i$ and $j$ up to $t_0$. 

Let $C$ be the set of cascades which start in $\mathcal{G}$ after time $t_0$. A snapshot of cascade $c\in C$ with a duration $t$ after its origination is characterized by a \textit{cascade graph} $g_c^t=(V_c^t,E_c^t)$, where $V_c^t$ is a subset of nodes in $V$ that have adopted the cascade $c$ within duration $t$ after its origination and $E_c^t = E\cap(V_c^t\times V_c^t)$, which is the set of edges in $E$ with both ends inside $V_c^t$. 

We consider the problem of predicting the \textbf{increment of the size} of cascade $c$ after a given time interval $\Delta t$, which is denoted as $\Delta s_c=|V_c^{t+\Delta t}|-|V_c^t|$. The cascade prediction can then be formulated as, given $\mathcal{G}$, $t$, $\Delta t$, and $\{(g_c^t,\Delta s_c)\}_{c\in C}$, finding an optimal mapping function $f$ that minimizes the following objective

\begin{equation}
	\mathcal{O} = {\frac{1}{|C|}\sum_c(f(g_c^t)- \Delta s_c)^2}
\end{equation}

In the definition, $t$ indicates the earliness of the prediction and $\Delta t$ indicates the horizon of the prediction. When $t$ is smaller, we are making predictions at the early stage of a cascade; when $\Delta t$ is larger, we are predicting the size of cascade that is closer to its final status. These scenarios are particularly valuable but inherently harder in reality. It is worth noting that we consider the social network structure $\mathcal{G}$ as static in the prediction task. While in reality the global network does change over time, we are doing this to control for the effect of cascades on the network structure in this study - new edges may form due to a particular information cascade. 

\subsection{DeepCas: the End-to-End Pipeline}

We propose an end-to-end neural network framework that takes as input of the cascade graph $g_c$ and predicts the increment of cascade size $\Delta s_c$. The framework (shown in figure \ref{fig:framework}) first samples node sequences from cascade graphs and then feeds the sequences into a gated recurrent neural network, where attention mechanisms are specifically designed to learn how to assemble sequences into ``documents,'' so that the future cascade size could be predicted. 
 
%\reminder{TODO: describe the general pipeline here. With the end-to-end pipeline, some of the other figures becomes redundant. Perhaps some of them can be omitted (e.g., Figure 3), or provide more details (in particular, the GRU and the attention). }

\subsection{Cascade Graph as Random Walk Paths}
\label{sec:cascdae_paths}
Given a cascade graph $g_c$, the first component in DeepCas generates an initial representation of $g_c$ using a set of node sequences. %We attempt to predict the future size of $g_c$, in other words, how influential a particular diffusion could be. The influence of a graph could be interpreted as the collected influence of the nodes in the cascade graph, which should not be treated as adopters but rather as ``propagators" of this diffusion. 

Naturally, the future size of a cascade highly depends on who the information ``propagators'' are, which are the nodes in the current cascade graph. Therefore, a straightforward way to represent a graph is to treat it as a bag of nodes. However, this method apparently ignores both local and global structural information in $g_c$, which have been proven to be critical in the prediction of diffusions~\cite{cheng2014can}. To remedy this issue, we sample from each graph a set of paths, instead of individual nodes. If we make a analogy between nodes and words, paths would be analogous to sentences, cascade graphs to documents, and a set of graphs to a  document collection. %These analogies mean that if we are able to sample paths from a graph, we could apply deep learning to graphs, in a very similar way as deep learning to documents.

\begin{figure}[ht]
\centering
\includegraphics[width=0.3\textwidth]{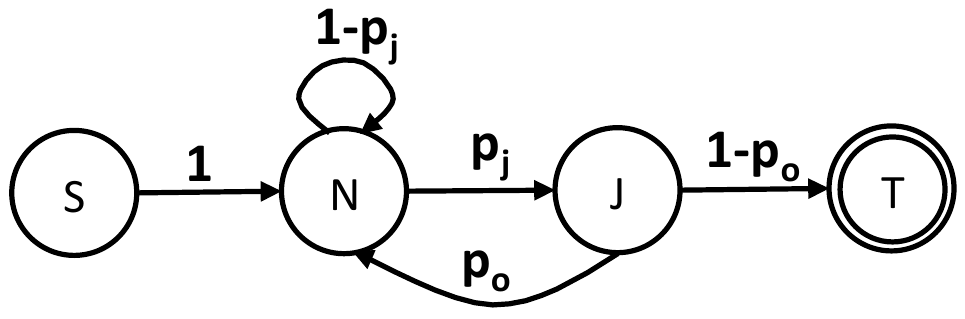}
\caption{The Markov chain of random walk.}
\label{fig:markov_chain}
\end{figure}

Similar to DeepWalk, the sampling process could be generalized as performing a random walk over a cascade graph $g_c$, the Markov chain of which is shown in Figure~\ref{fig:markov_chain}. The random walk for each diffusion graph starts from the starting state $S$, which is always followed by the state $N$, where the walker transits to a neighbor of the current node. With probability $1-p_j$, it goes on walking to the neighbor. With a jumping probability $p_j$, it jumps to an arbitrary node in the cascade graph, leading the walker to the jump state $J$. With continue probability $p_o$, it walks to a neighbor of the current node, thus going back to state $N$. With probability $1-p_o$, it goes to the terminal state $T$, terminating the entire random walk process. 

Suppose the walker is at state $N$ in the Markov chain and is currently visiting a node $v$, it follows a transition probability $p(u\in N_c(v)| v)$ to go to one of its outgoing neighbor $u\in N_c(v)$, where $N_c(v)$ denotes the set of $v$'s outgoing neighbors in diffusion graph $g_c$. There are multiple strategies for setting transition probabilities. Given a specific choice of scoring function $\mathrm{sc_t}(u)$ to transit to node $u$, the neighbor $u$ could be sampled in proportion to its score:
\begin{equation}
p(u\in N_c(v)| v) = \frac{\mathrm{sc_t}(u)+\alpha}{\sum_{s\in N_c(v)}(\mathrm{sc_t}(s)+\alpha)}
\label{equ:trans_prob}
\end{equation}
where $\alpha$ is a smoother. The scoring function $\mathrm{sc_t}(u)$ could be instantiated by but not limited to (1) $deg_c(u)$, the out-degree of node $u$ in $g_c$, (2) $deg_{\mathcal{G}}(u)$, the degree of $u$ in the global graph $\mathcal{G}$, or (3) $weight(v,u)$, the weight of the edge between the current node $v$ and its neighbor $u$. Likewise, when the walker is at state $J$ and is to select a node to jump to, the scoring function $\mathrm{sc_j}(u)$ could be set correspondingly. 
\begin{equation}
p(u) = \frac{\mathrm{sc_t}(u)+\alpha}{\sum_{s\in V_c}(\mathrm{sc_t}(s)+\alpha)}
\label{equ:jump_prob}
\end{equation}
 where $V_c$ is the node set of $g_c$, and $\mathrm{sc_t}(u)$ could be (1) $deg_c(u)$, (2) $deg_{\mathcal{G}}(u)$, or (3) $\sum_{s\in N_c(u)}weight(u,s)$. %, the weight sum of all outgoing edges of $u$.

\subsection{Sampling sequences from a graph}
\label{sec:sequence}
The probability $p_o$ of whether to perform another random jump or go to the terminal state essentially determines the  expected number of sampled sequences, while the probability $p_j$ of whether to perform a random jump or transit to neighbors corresponds to the sequence length. The two factors play a key role in determining the representations of cascade graphs. %, which is used to predict its influence power finally.

Naturally, different cascade graphs may require different parameters $p_o$ and $p_j$, as some are intrinsically more complex than others. Instead of fixing or manually tuning these two hyper-parameters, we propose to learn the two probabilities in an end-to-end manner by incorporating them to our deep learning framework. To do this, as Figure~\ref{fig:framework} (b) shows, we sample long enough sequences and sufficient number of sequences for all diffusion graphs. Denote $T$ the sampled sequence length, $K$ the sampled number of sequences, where $T$ and $K$ are the same for all diffusion graphs, we want to learn the actual length $t_c$ and the actual number of sequences $k_c$ we needed for each graph $g_c$, essentially a different parameterization of $p_o$ and $p_j$. 

Note that existing work of using random walk paths to represent graphs such as DeepWalk and Node2Vec use fixed, predefined $T$ and $K$. Automatically learning graph-specific path counts and lengths is a major technical contribution. We leave the learning of $t_c$ and $k_c$ to the next subsection.

%\begin{figure}[ht]
%\centering
%\includegraphics[width=0.45\textwidth]{figures/sampling.pdf}
%\caption{An example of sampling paths from a graph.}
%\label{fig:sample_path}
%\end{figure}

%% An example showing how to sample from a graph is displayed in Figure~\ref{fig:framework} (a) and (b). For each sequence, the starting node is sampled with probability according to Equation~\ref{equ:jump_prob}. Following this node, neighbors are sampled iteratively with probability according to Equation~\ref{equ:trans_prob}. The sampling of one sequence is stopped either when we reach the predefined length $T$, or when we reach a node without any outgoing neighbors. In this case, sequences with length less than $T$ nodes are padded by a special node `$+$'.  This process of sampling sequences continues until we sample $K$ sequences.

%A hard $l$ and $s$ are hard to learn, as they are not differentiable. To solve this problem, we use soft attention mechanism. We assume multinomial attention over $L$ nodes. For sequences, multinomial attention do not work well if we have thousands of sequences. Intuitively from the learning perspective, if we have already learned the representation of the diffusion well, we do not need to read more sequences. Therefore we use a geometric distribution over the attention of the sequences We divide sequences into mini-batches. Sequences within one batch share the same attention. The probability $p$ is conditioned on graph size.

\subsection{Neural Network Models}
\label{sec:DNN}
Once we have sampled $K$ sequences with $T$ nodes for each diffusion graph, any effective neural networks for sequences could be applied to the random walk paths in a similar way as to text documents. The output of the neural network gives us the hidden representation of individual sequences. Unlike documents whose sentences are already written, we have to learn how to ``assemble'' these individual sequences into a ``document,'' so that it can best represent the graph and predict its growth.

\paragraph*{\bf Node Embedding} Each node in a sequence is represented as a one-hot vector, $q \in \mathbf{R}^{N_{node}}$, where $N_{node}$ is the number of nodes in $\mathcal{G}$.  All nodes share an embedding matrix $A \in \mathbf{R}^{H \times N_{node}}$, which converts a node into its embedding vector $x=Aq, x\in \mathbf{R}^{H}$. 

%\begin{figure}[ht]
%\centering
%\includegraphics[width=0.45\textwidth]{figures/gru.pdf}
%\caption{x }
%\label{fig:gru}
%\end{figure}

\paragraph*{\bf GRU-based Sequence Encoding} The sampled sequences represent the flow of information of a specific diffusion item. To capture this information flow, we use a Gated Recurrent Unite (GRU)~\cite{hochreiter1997long}, a specific type of recurrent neural network (RNN). When applying GRU recursively to a sequence from left to right, the sequence representation will be more and more enriched by information from later nodes in this sequence, with the gating mechanism deciding the amount of new information to be added and the amount of history to be preserved, which simulates the process of information flow during a diffusion. Specifically, denote step $i$ the $i$-th node in a sequence, for each step $i$ with input node embedding $x_{i} \in \mathbf{R}^{H}$ and previous hidden state $h_{i-1} \in \mathbf{R}^{H}$ as inputs, GRU computes the updated hidden state $h_{i}=\textrm{GRU}(x_{i}, h_{i-1}), h_{i} \in \mathbf{R}^{H}$. %% by:
%% \begin{equation}
%% \begin{split}
%% u_{i} &= \sigma(W^{(u)}x_{i} + U^{(u)} h_{i-1} + b^{(u)}) \\
%% r_{i} &= \sigma(W^{(r)}x_{i} + U^{(r)} h_{i-1} + b^{(r)}) \\ 
%% \hat{h}_{i} &= \textrm{tanh}(W^{(h)}x_{i} + r_{i} \cdot U^{(h)} h_{i-1} + b^{(h)})\\
%% h_{i} &= u_{i} \cdot \hat{h}_{i-1} + (1-u_{i}) \cdot h_{i-1}
%% \end{split}
%% \end{equation}
%% where $\sigma(.)$ is the sigmoid activation function, $\cdot$ represents an element-wise product. $W^{(u)}$, $W^{(r)}$, $W^{(h)}$ $\in \mathbf{R}^{H \times d}$, $U^{(u)}$, $U^{(r)}$, $U^{(h)}$ $\in \mathbf{R}^{d \times d}$ are GRU parameters and $H$ is the hidden size. 

For now we have read the sequence from left to right. We could also read the sequence from right to left, so that earlier nodes in the sequence could be informed by which nodes have been affected by a cascading item passed from them. To this end, we adopt the bi-directional GRU, which applies a forward GRU that reads the sequence from left to right, and a backward GRU from right to left. We denote the forward GRU as $\textrm{GRU}_{fwd}$ and backward as $\textrm{GRU}_{bwd}$. As Figure~\ref{fig:framework} (c) shows, the presentation of the $i$-th node in $k$-th sequence, $\overleftrightarrow{h}_{i}^{k}  \in \mathbf{R}^{2H}$, is computed as the concatenation of the forward and backward hidden vectors. 
%% \begin{equation}
%% \begin{split}
%% \overrightarrow{h}_{i}^{k} & = \textrm{GRU}_{fwd}(x_{i}, \overrightarrow{h}_{i-1}^{k}) \\
%% \overleftarrow{h}_{i}^{k} & = \textrm{GRU}_{bwd}(x_{i}, \overleftarrow{h}_{i+1}^{k}) \\
%% \overleftrightarrow{h}_{i}^{k} & = \overrightarrow{h}_{i}^{k} \oplus \overleftarrow{h}_{i}^{k}  
%% \end{split}
%% \end{equation}
%% where $\oplus$ denotes the concatenation operation. 

% We refer the readers who are not familiar with GRU and RNNs to a general tutorial \cite{chung2014empirical}. 

\begin{figure}[ht]
\centering
\includegraphics[width=0.45\textwidth]{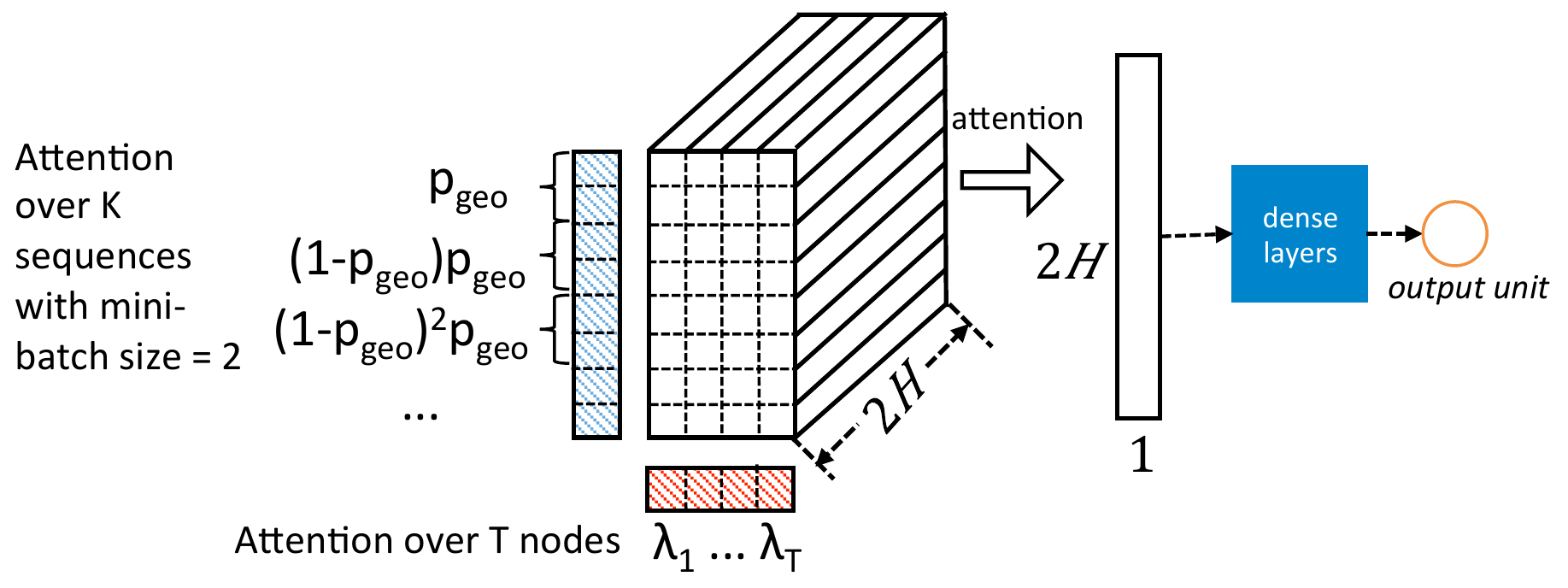}
\caption{Attention to assemble the representation of the graph.}
\label{fig:attention}
\end{figure}

\paragraph*{\bf From sequence to graph representation} Given a collection of sequence representations, where the $k$-th sequence with length $T$ is represented as $[\overleftrightarrow{h}_{1}^{k},$ $ ...,\overleftrightarrow{h}_{i}^{k},... $ $,\overleftrightarrow{h}_{T}^{k}]$, as displayed in Figure~\ref{fig:framework} (d), we attempt to learn the representation of the cascade graph as a whole, so that it best predicts its future size. Analogically, we are assembling a document (graph) from a large number of very long sentences. We do this by learning the number of sentences and length of sentences per document, through an \textit{attention} mechanism in deep learning. %We refer readers who are not familiar with attention in general to \footnote{\url{http://www.wildml.com/2016/01/attention-and-memory-in-deep-learning-and-nlp/}}.

In particular, the random walk on a graph terminates with probability $1-p_o$. From the learning perspective, we could learn the value of $p_o$ by examining whether the sampled number of sequences could represent the graph well, which in turn decides whether the prediction task is well performed. Intuitively, we could partition the sampled $K$ sequences into ``mini-batches.'' We want to read in more mini-batches until we could learn the graph well, simulating the action of jumping to the terminal state in the random walk. To implement this intuition, we assume a geometric distribution of \textit{attentions} over mini-batches. If sequences in the first mini-batch of cascade $g_c$ share attention weight $p_{geo}^c$, the next mini-batch will have attention $(1-p_{geo}^c)p_{geo}^c$, so on and so forth as Figure~\ref{fig:attention} shows. In theory, if we sample infinite number of sequences with the geometric distribution so that $K\to\infty$, the number of expected mini-batches to learn will be $1/p_{geo}^c$. %In practice, we sample large enough number of sequences to approximate this expectation. 
With this expectation, learning the parameter $p_{geo}^c$ could help us decide how many sequences to read in. Note that the degree of freedom is too high if we fit a free parameter $p_{geo}^c$ per cascade. Instead, we rely on an observation that the number of sequences we need to represent a cascade graph is correlated with its size. Therefore, we condition $p_{geo}^c$ on the size of graph $\mathrm{sz}(g_c)$, more specifically $\floor{\log_2(\mathrm{sz}(g_c)+1)}$. %, where $\floor{\cdot}$ takes the floor of a floating number. 
As a result, $p_{geo}^c$ is replaced with $p_{geo}^{\floor{\log_2(\mathrm{sz}(g_c)+1)}}$.

We could apply similar procedure to learn sequence length. In practice, we found that the standard multinomial distribution of attentions already work well. So we simply assume multinomial distribution $ \lambda_{1}, ..., \lambda_{T}$ over $T$ nodes so that $\sum_i(\lambda_{i})=1$, where $\{\lambda_{i}\}$ are shared across all cascade graphs.

To sum up and to give a mathematical representation, suppose the mini-batch size is $B$ sequences, then the $k$-th sequence will fall into $(\floor{k/B}+1)$-th mini-batch, the attention mechanism then outputs the representation for graph $g_c$, a vector of length $2H$:
%of the $K$ sampled sequences with length $T$, we adopt an attention mechanism to choose which vectors to focus on. The attention mechanism then produces a low-dimensional vector representation for a graph with node information. Specifically, we adopted a two level attention mechanism over both sequences and nodes. The attention over sequences is conditioned on the size of the graph $a_{1}(G), ..., a_{K}(G)$ and the attention over nodes follows multinomial distribution $ \lambda_{1}, ..., \lambda_{T}$.  $h(\mathbf{G})$:
\begin{equation}
\begin{split}
h(g_c) = \sum_{k=1}^{K} \sum_{i=1}^{T}\left( (1-a_c)^{\floor{k/B}}a_c\right)\lambda_{i} \overleftrightarrow{h}_{i}^{t},
\end{split}
\end{equation}
where the first term corresponds to the attention over sequences with geometric distribution, and $a_c = p_{geo}^{\floor{\log_2(\mathrm{sz}(g_c)+1)}}$. Both $a_c$ and $\lambda_i$ are learned through the deep learning process.

\paragraph*{\bf Output module} Our output module consists of a fully connected layer with one final output unit: $f(g_c) = \textrm{MLP}(h(g_c))$, where MLP stands for a multi-layer perceptron. 

\section{Experiment setup}
\label{sec:exp_setup}

We present comprehensive empirical experiments using real world data sets to evaluate the performance of DeepCas. 

\subsection{Data Sets}
Most existing work evaluates their methods of predicting diffusions on a single social network data set (e.g., ~\cite{cheng2014can, cui2013cascading, guo2015toward}. We add another completely different, publicly available data set to demonstrate the effectiveness and generalizability of DeepCas and to allow readers to reproduce our results. 

One of the scenario is the cascade of Tweets on Twitter. Following the practice in existing work~\cite{ratkiewicz2011truthy}, we collect the \textsc{Twitter} data set which contains the cascades of Tweets (i.e., through retweeting) in June, 2016 from the official Decahose API (10\% sample of the entire Tweet stream). All original English tweets that are published from June 1 to June 15 and retweeted at least once in 10 days are used for training. Those with only one retweets are downsampled to 5\%. Cascades originated on June 16 are used for validation, and cascades originated from June 17 to June 20 are used for testing. 
%The set of included tweet users are formed as follows. %For all tweets in training, we collected all the authors and retweeters in June, with the number of users reported in Table~\ref{tab:datasets}. 
A cascade contains the authors of the original Tweet and its retweets. 
%When constructing cascade graphs, we only consider the cascade among these users, which is the best available practice in literature~\cite{bourigault2016representation,cui2013cascading}. \reminder{TODO: will need to revisit this paragraph.}

We construct the global social network $\mathcal{G}$ using the same Tweet stream in April and May 2016. 
As the follower/followee relations are not available in the data and Twitter does not disclose the retweet paths, we follow existing work~\cite{ratkiewicz2011truthy} and %buil edges between users based on the actual flow of information. An edge is 
draw an edge from Twitter user A to B if either B retweeted a message of A or A mentioned B in a Tweet. Comparing to a follower/followee network, this network structure accumulates all information cascades and reflects the truly active connections between Twitter users. %, which could be even better for cascade prediction. 
We weigh an edge based on the number of retweeting/mentioning events between the two users.
To construct cascade graphs, we choose $t$, the duration of cascade since the original Tweet was posted, from a range of $t=1,3,5$ days. We compute the increment of cascade size after $t$ for the next $\Delta t$ days, where $\Delta t=1,3,5$ days. The combination of $t$ and $\Delta t$ yields a total of $3 \times 3 = 9$ configurations.

In the second scenario, we evaluate the prediction of the cascades of scientific papers. We collect the \textsc{AMiner} data set using the DBLP citation network released by ArnetMiner \footnote{\url{https://aminer.org/citation}, DBLP-Citation-network V8, retrieved in August 2016.}. We construct the global network $\mathcal{G}$ based on citations between 1992 and 2002. That is, an edge draws from node A to B if author A is ever cited by B (which indicates that B might have found a reference from reading A's papers). A cascade of a given paper thus involves all authors who have written or cited that paper. Papers published between 2003 and 2007 are included in the training set. Papers published in 2008 and 2009 are used for validation and testing, respectively.  For the earliness and horizon of predictions, we set $t=1,2,3$ years and $\Delta t=1,2,3$ years respectively.

In both scenarios, we notice that the growth of all the cascades follows a power-law distribution, where a large number of cascades did not grow at all after $t$. Therefore we downsample 50\% graphs with zero growth (to the numbers shown in Table~\ref{tab:datasets}) and apply a logarithm transformation of the outcome variable (increment of cascade size), following existing literature~\cite{kupavskii2012prediction,tsur2012s}.

\begin{table}[h!]
\caption{Statistics of the data sets.}
\small 
\begin{center}
\setlength{\tabcolsep}{3pt}
\begin{tabular}{c|c|c|c|c||c|c|c} 
\hline 
 &  Set & \multicolumn{3}{c||}{\textsc{Twitter}}  & \multicolumn{3}{c}{\textsc{AMiner}}  \\
\hline 
\# nodes in $\mathcal{G} $ & All & \multicolumn{3}{c||}{354,634}  & \multicolumn{3}{c}{131,415}   \\
\hline 
\# edges in $\mathcal{G} $ & All &  \multicolumn{3}{c||}{27,929,863}  & \multicolumn{3}{c}{842,542}  \\
\hline 
t &  & 1 day & 3 days & 5 days & 1 year & 2 years & 3 years \\
\hline 
 & train & 25,720 & 26,621 & 26,871 & 3,044 & 17,023 & 34,347 \\
\# cascades & val & 1,540 & 1,563 & 1,574 & 509 & 3,665 & 7,428 \\
 & test & 6,574 & 6,656 & 6,663 & 517 & 3,512 & 7,337 \\
\hline 
 & train & 26.2 & 34.9 & 39.1 & 16.4 & 16.8 & 19.7 \\
Avg. nodes  & val & 46.1 & 62.1 & 69.7 & 10.6 & 13.6 & 17.2 \\
per $g_c$ & test & 50.8 & 65.8 & 72.8 & 8.8 & 12.6 & 16.2 \\
\hline 
 & train & 99.0 & 153.8 & 188.3 & 56.8 & 54.9 & 68.5 \\
Avg. edges  & val & 167.0 & 241.4 & 296.5 & 29.5 & 40.9 & 55.3 \\
per $g_c$ & test & 162.3 & 242.2 & 289.0 & 22.6 & 32.9 & 44.5 \\
\hline 
%% & train & 1.1 & 0.6 & 0.5 & 1.8 & 2.2 & 2.0 \\
%% Avg. scaled  & val & 1.4 & 0.9 & 0.7 & 1.9 & 2.0 & 1.8 \\
%% growth $\Delta t_1$ & test & 1.3 & 0.8 & 0.7 & 1.7 & 2.0 & 1.8 \\
%% \hline 
%%  & train & 1.6 & 1.2 & 1.1 & 2.5 & 3.0 & 2.8 \\
%% Avg. scaled  & val & 2.1 & 1.6 & 1.3 & 2.7 & 2.8 & 2.6 \\
%% growth $\Delta t_2$ & test & 1.9 & 1.4 & 1.3 & 2.4 & 2.8 & 2.5 \\
%% \hline 
%% & train & 1.9 & 1.5 & 1.3 & 3.0 & 3.5 & 3.2 \\
%% Avg. scaled  & val & 2.4 & 1.9 & 1.7 & 3.1 & 3.3 & 3.0 \\
%% growth $\Delta t_3$ & test & 2.2 & 1.8 & 1.6 & 2.9 & 3.2 & 2.8 \\
%% \hline 
\end{tabular}
%% \flushleft\textit{Avg. scaled growth} scales label $y$ to $\log_2(y+1)$~\cite{kupavskii2012prediction,tsur2012s}.\\
%% \textit{$\Delta t_1$, $\Delta t_2$, $\Delta t_3$} are respectively 1,3,5 days for tweets, and 1,2,3 years for papers.\\
\end{center}
\label{tab:datasets}
\end{table}

%% As we observe from Table~\ref{tab:datasets}, the validation and test sets of \textsc{Twitter} have larger number of nodes and edges on average than the training set, while \textsc{AMiner} appears to be the opposite. \reminder{I don't quite understand the followign sentence:} We remove a cascade from the data set if none of the users in train appear in the diffusion graph, thus diffusion with smaller sizes are more likely to be removed. As a result, the validation and test sets of Tweets contain more graphs of larger sizes. As for the academic data set, since the training period spans five years, many authors in this period might have became inactive in later stage. Consequently, there will be fewer active authors in the testing period, leading to the shrinkage of the observed cascades.

\subsection{Evaluation Metric}
We use the mean squared error (MSE) to evaluate the accuracy of predictions, which is a common choice for regression tasks and used in previous work of cascade prediction~\cite{tsur2012s,yu2015micro,kupavskii2012prediction}. %% Denote $\hat{y}$ a prediction value, and $y$ the ground truth value, the MSE is:
%% \begin{equation}
%% \textrm{MSE}=\frac{1}{n}\sum_{i=1}^n(\hat{y}_{i} - y_i)^2
%% \end{equation}
As noted in Section~\ref{sec:problem}, %% $y_i$ in above equation is 
we predict a scaled version of the actual increment of the cascade size, i.e., $y_i=\log_2(\Delta s_i + 1)$.

\subsection{Baseline methods}
We compare DeepCas with a set of strong baselines, including feature-based methods used for cascade prediction, methods based on nodes embeddings, and alternative deep learning methods to learn graph representations. 

\textbf{Features-}. We include all structural features that could be generalized across data sets from recent studies of cascade prediction~\cite{cheng2014can,guo2015toward,cui2013cascading,ugander2012structural}. These features include:

\textit{Centrality and Density}. Degree of nodes in the cascade graph $g$ and the global network $\mathcal{G}$, average and 90th percentile of the local and global degrees of nodes in $g$, number of leaf nodes in $g$, edge density of $g$, and the number of nodes and edges in the \textit{frontier graph} of the cascade, which is composed of nodes that are not in $g$ but are neighbors of nodes in $g$.

\textit{Node Identity.} The presence of node ids in $g$ is used as features.

\textit{Communities}. From both the cascade graph and the frontier graph, we compute the number of communities ~\cite{blondel2008fast}, the overlap of communities, and Gini impurity of communities~\cite{guo2015toward}.

\textit{Substructures.} We count the frequency of k-node substructures ($k\leq 4$)~\cite{ugander2012structural}. These include nodes ($k=1$), edges ($k=2$), triads (e.g., the number of closed and open triangles) and quads from both the cascade graph and the frontier graph.

\textbf{-linear} and \textbf{-deep}. Once the cascade is represented as a set of features above, they are blended together using linear regression (denoted as \textbf{Features-linear}) with L2 regularization, as other linear regressors such as SVR empirically perform worth on our task. %We append  to each method to indicate usage of linear regression. 
To obtain an even stronger baseline, we feed the feature vectors to MLP (denoted as ~\textbf{Features-deep}).

\textbf{OSLOR} selects important nodes as sensors, and predict the outbreaks based on the cascading behaviors of these sensors~\cite{cui2013cascading}.

\textbf{Node2vec}~\cite{grovernode2vec} is selected as a representative of node embedding methods. Node2vec is a generalization of DeepWalk~\cite{perozzi2014deepwalk}, which is reported to be outperforming alternative methods such as DeepWalk and LINE~\cite{tang2015line}. We generate walks from two sources: (1) the set of cascade graphs $\{g\}$ (2) the global network $\mathcal{G}$. The two sources lead to two embedding vectors per node, which are concatenated to form the final embedding of each node. The average of embeddings of all nodes in a cascade graph is fed through MLP to make the prediction.

\textbf{Embedded-IC}~\cite{bourigault2016representation} %Based on the \reminder{Independent Cascade model}, embedded-IC 
represents nodes by two types of embeddings: as a sender or as a receiver. For prediction, the original paper used Monte-Carlo simulations to estimate infections probabilities of each individual user. To predict cascade size, we experiment with two settings: (1) learn a linear mapping function between the number of infected users and the cascade size; (2) follow the setting of Node2Vec by using the average of embeddings of all nodes in the cascade graph, which is then piped through MLP. We find that the second setting empirically performs better than the first one. We therefore report the performance of the latter.

\textbf{PSCN} applies convolutional neural networks (CNN) to locally connected regions from graphs~\cite{niepert2016learning}. We apply PSCN to both the diffusion graphs and the frontier graphs. The last hidden layer of the cascade graph and that of the frontier graph are concatenated to make the final prediction.

\textbf{Graph kernels}. There are a set of state-of-the-art graph kernels~\cite{niepert2016learning}: the shortest-path kernel (SP)~\cite{borgwardt2005shortest}, the random walk kernel (RW)~\cite{gartner2003graph}, and the Weisfeiler-Lehman subtree kernel (WL) \cite{shervashidze2011weisfeiler}. The RW kernel and the SP kernel are too computationally inefficient, which did not complete after 10 days for a single data set in our experiment. We therefore exclude them from the comparison, a decision also made by in~\cite{niepert2016learning,yanardag2015structural}. For the WL kernel, we experiment with two settings: \textbf{WL-degree}, where node degree is used as the node attribute to build subgraphs for each cascade and frontier graph; \textbf{WL-id}, where node id is used as the attribute. The second setting is to test whether node identity information could be incorporated into graph kernel methods.

% TODO: Hyper-parameters
\textbf{Hyper-parameters}. All together we have 8 baselines. All their hyper-parameters are tuned to obtain the best results on validation set for each configuration (9 in total) of each data set. For linear regression, we chose the L2-coefficient from $\{1, 0.5,$ $0.1,$ $0.05, ..., 10^{-8}\}$. For neural network regression, the initial learning rate is selected from $\{0.1,$ $0.05,$ $0.01,$ $...,$ $10^{-4}\}$, the number of hidden layers from $\{1, 2, ..., 4\}$, the hidden layer size from $\{32, 64, ..., 1024\}$, and L1- and L2-coefficient both from $\{1, 0.5,$ $0.1,$ $0.05, ..., 10^{-8}\}$. Following~\cite{niepert2016learning} for PSCN, the width is set to the average number of nodes, and the receptive field size is chosen between 5 and 10. The height parameter of WL is chosen from $\{2, 3, 4\}$. The candidate embedding size set is selected from $\{50,$ $100,$ $200,$ $300\}$ for all methods that learn embeddings for nodes. For node2vec, we follow~\cite{grovernode2vec}, $p$, $q$ are selected from $\{0.25, 0.50, $ $1, $ $2, 4\}$, the length of walk is chosen from $\{10, 25, 50, $ $75, $ $100\}$, and the number of walks per node is chosen from $\{5, 10, $ $15, $ $20\}$.

\subsection{DeepCas and the Variants}
We compare a few variants of DeepCas with the 8 baselines. We sample $K=200$ paths each with length $T=10$ from the cascade graph without tuning the parameters. As described in Section~\ref{sec:sequence} and ~\ref{sec:DNN}, the attention mechanism will automatically decide when and where to stop using the sequences. The mini-batch size is set to 5. The smoother $\alpha$ is set to 0.01. The embedding sizes for the \textsc{Twitter} and \textsc{AMiner} data set are set to 150 and 50 respectively. The embeddings are initialized by concatenating embedding learned by Node2Vec from both all diffusion graphs $\{g\}$ in training set and the global network $\mathcal{G}$. The node2vec hyper-parameters $p$ and $q$ are simply set to 1.

We use \textbf{DeepCas-edge}, \textbf{DeepCas-deg}, and \textbf{DeepCas-DEG} to denote three version of DeepCas, which randomly walk with transition probabilities proportional to edge weights, node degree in the cascade graph, and node degree in the global network. For comparison, we also include three simplified versions of DeepCas:

%\subsubsection{Variants of methods}

\textbf{GRU-bag} represents a cascade graph as a bag of nodes and feeds them through our GRU model. This is similar to setting the length of random walk paths to 1, which examines whether sequential information is important for cascade prediction.

\textbf{GRU-fixed} uses a fixed path length $t$ and a fixed number of sequences $k$, without using the attention mechanism to learn them adaptively. Hyper-parameters $t$ and $k$ are tuned to optimal on the validation sets, the values of which are selected from $\{2, 3, 5, $ $ 7, $ $ 10\}$ and from $\{50, 100, $ $ 150, $ $ 200\}$, respectively.

\textbf{GRU-root} uses the attention mechanism, but starts sampling a random walk path only from roots, which are nodes who started the diffusion. If there are multiple roots, we take turns to sample from them. This examines whether it is important to perform random jumps in the walks over the graph.

%Below we present the results of the experiments. 

\section{Experiment results}
\label{sec:exp}

%In this section, we present the results of the experiments as designed in Section 4. 

\subsection{Overall performance}
\begin{table*}[h!]
\caption{Performance measured by MSE (the lower the better), where original label $\Delta s$ is scaled to $y = \log_2(\Delta s+1)$.}
\begin{center}
\small
\subcaption*{(a) \textsc{Twitter}}
\begin{tabular}{c|c|c|c||c|c|c||c|c|c} 
\hline 
$t$ &  \multicolumn{3}{c||}{1 day}   &   \multicolumn{3}{c||}{3 days}  & \multicolumn{3}{c}{5 days}   \\
\hline 
$\Delta t$ & 1 day & 3 days & 5 days & 1 day & 3 days & 5 days & 1 day & 3 days & 5 days \\
\hline 
Features-deep & 1.644  & 2.253  & 2.432  & 1.116  & 1.687  & 2.133  & 0.884  & 1.406  & 1.492  \\
\hline 
Features-linear & $1.665^{**}$ & 2.256  & $2.464^{**}$ & 1.123  & $1.706^{*}$ & 2.137  & 0.885  & $1.425^{*}$ & 1.505  \\
\hline 
OSLOR & $1.791^{***}$ & $2.485^{***}$ & $2.606^{***}$ & $1.179^{***}$ & $1.875^{***}$ & $2.181^{***}$ & $0.990^{***}$ & $1.539^{***}$ & $1.778^{***}$ \\
\hline 
node2vec & $1.759^{***}$ & $2.384^{***}$ & $2.562^{***}$ & $1.145^{**}$ & $1.760^{***}$ & 2.143  & 0.895  & $1.460^{***}$ & $1.544^{***}$ \\
\hline 
Embedded-IC & $2.079^{***}$ & $2.706^{***}$ & $2.944^{***}$ & $1.277^{***}$ & $2.072^{***}$ & $2.316^{***}$ & $1.012^{***}$ & $1.743^{***}$ & $1.955^{***}$ \\
\hline 
PSCN & $1.735^{***}$ & $2.862^{***}$ & $2.911^{***}$ & $1.134^{*}$ & $1.784^{***}$ & $2.411^{***}$ & 0.893  & $1.461^{***}$ & $1.566^{***}$ \\
\hline 
WL-degree & $1.778^{***}$ & $2.568^{***}$ & $2.691^{***}$ & $1.177^{***}$ & $1.890^{***}$ & $2.205^{***}$ & $0.939^{***}$ & $1.568^{***}$ & $1.825^{***}$ \\
\hline 
WL-id & $1.805^{***}$ & $2.611^{***}$ & $2.745^{***}$ & $1.357^{***}$ & $1.967^{***}$ & $2.197^{***}$ & $0.945^{***}$ & $1.602^{***}$ & $1.853^{***}$ \\
\hline 
\multicolumn{1}{l|}{\tiny Proposed methods}&&&&&&&&&\\
GRU-bag & $1.769^{***}$ & $2.374^{***}$ & $2.565^{***}$ & $1.172^{***}$ & $1.822^{***}$ & 2.159  & $0.932^{***}$ & $1.472^{***}$ & $1.594^{***}$ \\
\hline 
GRU-fixed & $1.606^{**}$ & $2.149^{***}$ & $2.286^{***}$ & $1.132^{*}$ & 1.675  & $1.825^{***}$ & 0.891  & $1.376^{***}$ & $1.513^{*}$ \\
\hline 
GRU-root & $1.572^{***}$ & $2.202^{**}$ & $2.147^{***}$ & 1.097  & $1.726^{***}$ & $1.762^{***}$ & 0.874  & 1.406  & 1.489  \\
\hline 
DeepCas-edge & \boldmath$1.480^{***}$ & $1.997^{***}$ & $2.074^{***}$ & $1.013^{***}$ &\boldmath $1.567^{***}$ & $1.735^{***}$ & $0.854^{***}$ & \boldmath$1.322^{***}$ & $1.422^{***}$ \\
\hline 
DeepCas-deg & $1.492^{***}$ & \boldmath$1.933^{***}$ & \boldmath$2.033^{***}$ & $1.039^{***}$ & $1.597^{***}$ & \boldmath$1.707^{***}$ & $0.854^{***}$ & $1.330^{***}$ & \boldmath$1.412^{***}$ \\
\hline 
DeepCas-DEG & $1.487^{***}$ & $2.124^{***}$ & $2.081^{***}$ & \boldmath$1.012^{***}$ & $1.644^{***}$ & $1.724^{***}$ & \boldmath$0.849^{***}$ & 1.409  & $1.457^{***}$ \\
\hline 
\end{tabular}

\subcaption*{(b) \textsc{AMiner}}
\begin{tabular}{c|c|c|c||c|c|c||c|c|c} 
\hline 
$t$ &  \multicolumn{3}{c||}{1 year}   &   \multicolumn{3}{c||}{2 years}  & \multicolumn{3}{c}{3 years}   \\
\hline 
$\Delta t$ & 1 year & 2 years & 3 years & 1 year & 2 years & 3 years & 1 year & 2 years & 3 years \\
\hline 
Features-deep & 1.748  & 2.148  & 2.199  & 1.686  & 1.876  & 1.954  & 1.504  & 1.617  & 1.686  \\
\hline 
Features-linear & 1.737  & 2.145  & 2.205  & 1.690  & 1.887  & 1.964  & $1.529^{**}$ & 1.626  & 1.697  \\
\hline 
OSLOR & 1.768  & 2.173  & 2.225  & $1.897^{***}$ & $1.964^{***}$ & $2.057^{***}$ & $1.706^{***}$ & $1.738^{***}$ & $1.871^{***}$ \\
\hline 
node2vec & 1.743  & 2.153  & 2.209  & 1.702  & $1.921^{***}$ & $1.999^{***}$ & $1.563^{***}$ & $1.708^{***}$ & $1.816^{***}$ \\
\hline 
Embedded-IC & $2.117^{***}$ & $2.576^{***}$ & $2.751^{***}$ & $2.113^{***}$ & $2.429^{***}$ & $2.551^{***}$ & $1.947^{***}$ & $2.183^{***}$ & $2.285^{***}$ \\
\hline 
PSCN & $1.880^{**}$ & $2.332^{***}$ & $2.424^{***}$ & $1.853^{***}$ & $2.164^{***}$ & $2.092^{***}$ & $1.770^{***}$ & $1.822^{***}$ & $1.857^{***}$ \\
\hline 
WL-degree & 1.742  & $2.234^{*}$ & $2.350^{**}$ & 1.780  & $2.037^{***}$ & $2.079^{***}$ & $1.586^{***}$ & $1.762^{***}$ & $1.864^{***}$ \\
\hline 
WL-id & $2.566^{***}$ & $2.779^{***}$ & $2.900^{***}$ & $2.100^{***}$ & $2.259^{***}$ & $2.297^{***}$ & $2.029^{***}$ & $2.076^{***}$ & $2.086^{***}$ \\
\hline 
\multicolumn{1}{l|}{\tiny Proposed methods}&&&&&&&&&\\
GRU-bag & 1.783  & 2.217  & 2.242  & $1.712^{*}$ & $1.982^{***}$ & $1.988^{**}$ & $1.614^{***}$ & $1.743^{***}$ & $1.856^{***}$ \\
\hline 
GRU-fixed & 1.703  & 2.064  & 2.151  & $1.569^{***}$ & $1.735^{***}$ & $1.805^{***}$ & $1.430^{***}$ & $1.537^{***}$ & $1.564^{***}$ \\
\hline 
GRU-root & $1.816^{*}$ & $2.222^{*}$ & $2.331^{**}$ & $1.890^{***}$ & $1.972^{***}$ & $2.146^{***}$ & $1.660^{***}$ & $1.778^{***}$ & $1.813^{***}$ \\
\hline 
DeepCas-edge & \boldmath$1.668^{*}$ & \boldmath$2.016^{**}$ & \boldmath$2.084^{*}$ & $1.545^{***}$ & \boldmath$1.693^{***}$ & $1.799^{***}$ & \boldmath$1.402^{***}$ & $1.477^{***}$ & $1.548^{***}$ \\
\hline 
DeepCas-deg & $1.684^{*}$ & $2.043^{*}$ & $2.113^{*}$ & $1.544^{***}$ & $1.716^{***}$ & \boldmath$1.792^{***}$ & $1.407^{***}$ & \boldmath$1.469^{***}$ & $1.545^{***}$ \\
\hline 
DeepCas-DEG & $1.685^{*}$ & $2.036^{*}$ & $2.107^{*}$ & \boldmath$1.540^{***}$ & $1.700^{***}$ & $1.788^{***}$ & $1.404^{***}$ & $1.480^{***}$ & \boldmath$1.527^{***}$ \\
\hline 
\end{tabular}

``***(**)" means the result is significantly better or worse over \textit{Features-deep} according to paired t-test test at level 0.01(0.1).
\end{center}
\label{tab:mse_results}
\end{table*}

The overall performance of all competing methods across data sets are displayed in Table~\ref{tab:mse_results}. The last three rows of each table show the performance of the complete versions of our methods, which outperform all eight baseline methods with a statistically significant drop of MSE. Please note that the numbers in Table~\ref{tab:mse_results} are errors of log-transformed outcomes. If we translate them back to raw sizes, the numerical differences between the methods would look larger.

%\reminder{TODO: how large? Skipped as experiment results are not easy to explain}.

%\reminder{TODO: need to add more discussion about the results. Features-linear outperforms Features-deep is worth mentioning. That shows deep learning not always better than linear if we have already found good features. End-to-end is the key. Also worth comparing the error over different configurations. }

%Though feature based methods are able to incorporate a variety of hand-crafted features, without learning end-to-end, information could easily be lost. 

The difference between Features-deep and Features-linear is intriguing, which shows that deep learning does not always perform better than linear methods if we have already found a set of good features. It is more important to learn  end-to-end from the data.

Node2Vec and Embedded-IC do not perform well in cascade prediction. Taking the average of node embeddings as the graph representation is not as informative as representing the graph as a set of paths, even if the node embeddings are also fed into a deep neural net to make predictions. 
By comparing WL-degree and WL-id, we can see that it is hard for graph kernels to incorporate node identities. Simply using identities as node labels degenerates performance. This is because graph kernels rely on node labels to compute similarity between graphs. Using node id to measure similarity could cause serious sparsity problem.

The three simplified versions of DeepCas, GRU-bag, GRU-fixed, and GRU-root all lead to certain degradation of performance, comparing to the three DeepCas models. This empirically proves the effectiveness of the three important components of DeepCas. First, sampling a set of paths to represent a graph instead of averaging the representations of nodes is critical, as it facilitates the learning of structural information. Second, learning the random walks by adaptively deciding when to stop sampling from a particular path and when to stop sampling more paths is more effective than using a fixed number of fixed-length paths (which is what DeepWalk does). The suitable numbers and lengths might be associated with the complexity and the influence power of a cascade graph. If a cascade graph is more complex and more ``influential,'' it needs more paths and longer paths to represent its power. %DeepCas is able to adapt this automatically without explicitly measuring such factors. 
Third, sampling paths only from the root is not adequate (which is what most generative models do). Randomly jumping to other nodes could make the graph representation carry more information of the cascade structure and handle missing data. In a way, this is related to the ``mysterious'' feature used in Cheng et al. \cite{cheng2014can}, i.e., whether some early adopters are not directly connected to the root. 

Comparing the performance of using different $t$ and $\Delta t$, we see a general pattern that can be applied to all methods: the larger the earliness $t$ is, the easier to make a good prediction. This is because longer $t$ makes more information available. While for $\Delta t$, it is the opposite, as it is always harder to make long-term predictions.

%\subsection{Computational Cost}
%The computational cost of DeepCas can be divided into two parts: the generation of random walk paths and the training of deep learning models. 
Training DeepCas is quite efficient. On a machine with 2.40 GHz CPU, 120G RAM and a single Titan X GPU, it takes less than 20 minutes to generate random walk paths for a complete data set and less than 10 minutes to train the deep neural network. %on a data set is very fast, which is converged in less than 10 minutes.

%\subsection{Feature Analysis}
%\begin{figure}[h]
%\centering
%\includegraphics[width=0.4\textwidth]{figs/papers_vis.pdf}
%\caption{Feature visualization.}
%\label{fig:feature_analysis}
%\end{figure}

%The statistics of graphs where either 
We also investigate cascades for which DeepCas makes more mistakes than the baselines, and also the other way around. DeepCas tend to perform better on larger and denser graphs. These structures are more complex and harder to be represented as a bag of hand-crafted features. An end-to-end predictor without explicit feature design works very well in these cases. %or the baseline significantly outperforms the other are in general higher than the average statistics of each data set. This could result form the skewed distribution of the data set -- a large number of graphs are of smaller size, leading to more training instances of small graphs. 
%Both methods perform reasonably well on smaller graphs. 
For the sake of space, we omit the detailed statistics here. %% so the variance of performance is small.

\subsection{Interpreting the Representations}
We have empirically shown that DeepCas could learn the representation of cascade graphs that incorporates both structures and node identities. Qualitatively, we have not assessed what the learned representation actually captures from these information. Indeed, one concern of applying deep learning to particular domains is that the models are black-boxes and not easy to interpret. For us, it is intriguing to know whether the learned representation corresponds to well-known network properties and structural patterns in literature. 

\begin{figure}[h!]
\begin{minipage}[t]{0.44\textwidth}
\centering
\begin{minipage}[t]{0.44\textwidth}
\centering
\includegraphics[width=0.90\textwidth]{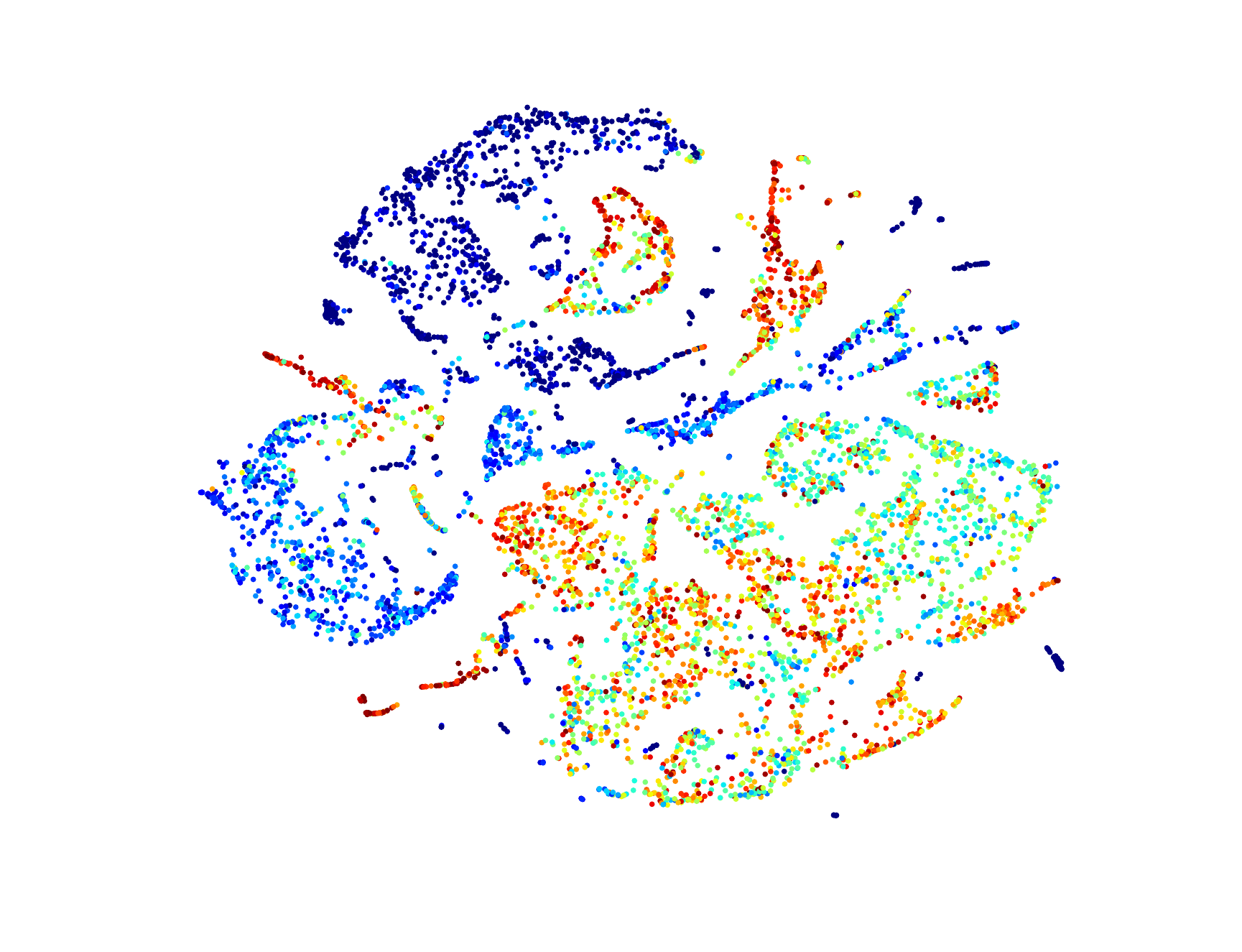}\\
(a) \# closed triangles.
\end{minipage}
\begin{minipage}[t]{0.44\textwidth}
\centering
\includegraphics[width=0.98\textwidth]{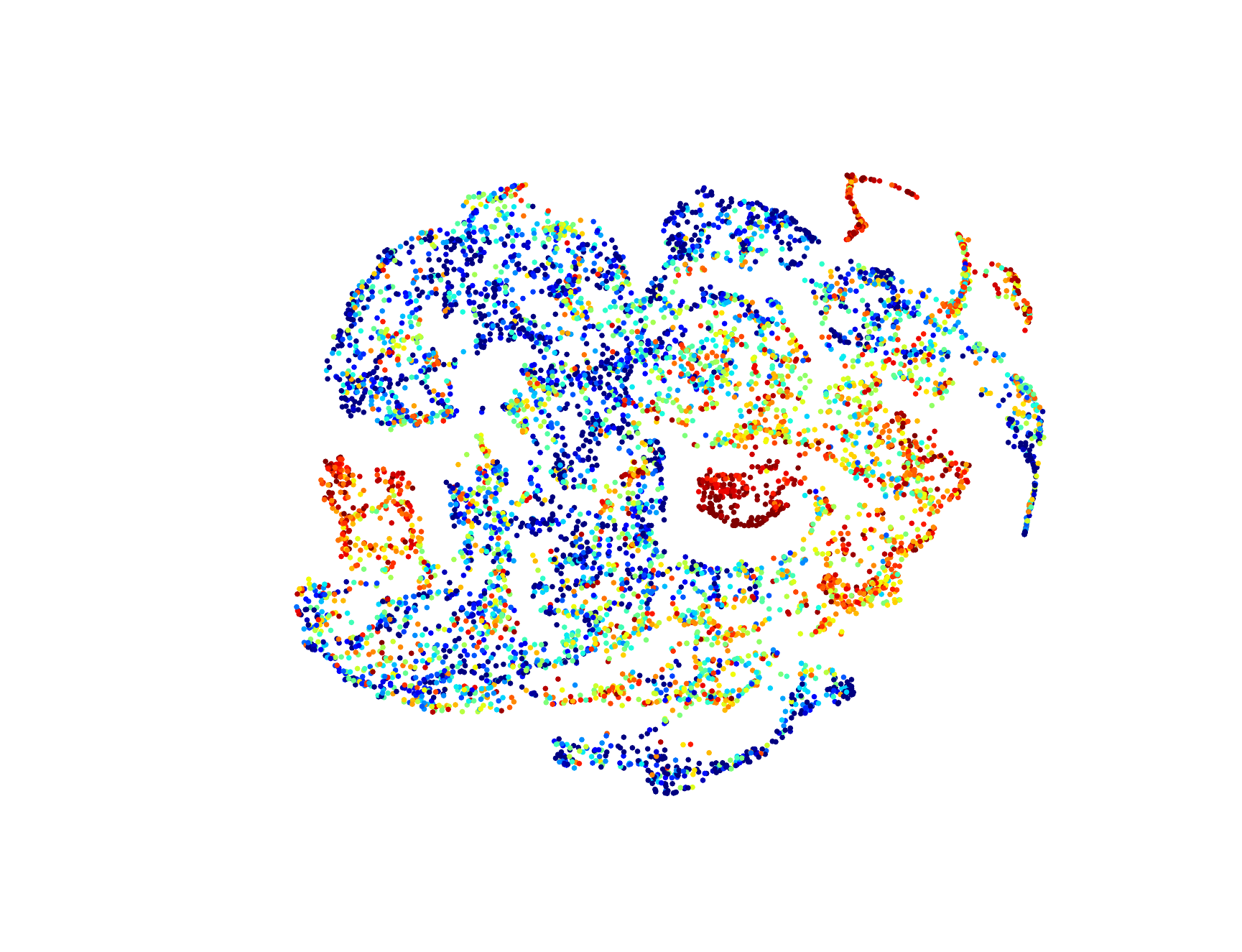}\\
(b) \# open triangles.
\end{minipage}
\begin{minipage}[t]{0.06\textwidth}
\centering
\includegraphics[width=0.9\textwidth]{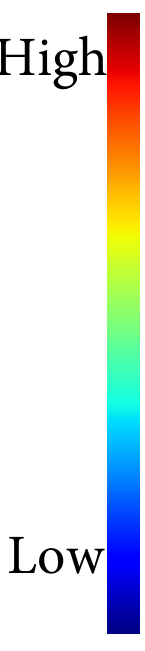}\\
\end{minipage}
\end{minipage}
\begin{minipage}[t]{0.44\textwidth}
\centering
\begin{minipage}[t]{0.44\textwidth}
\centering
\includegraphics[width=0.90\textwidth]{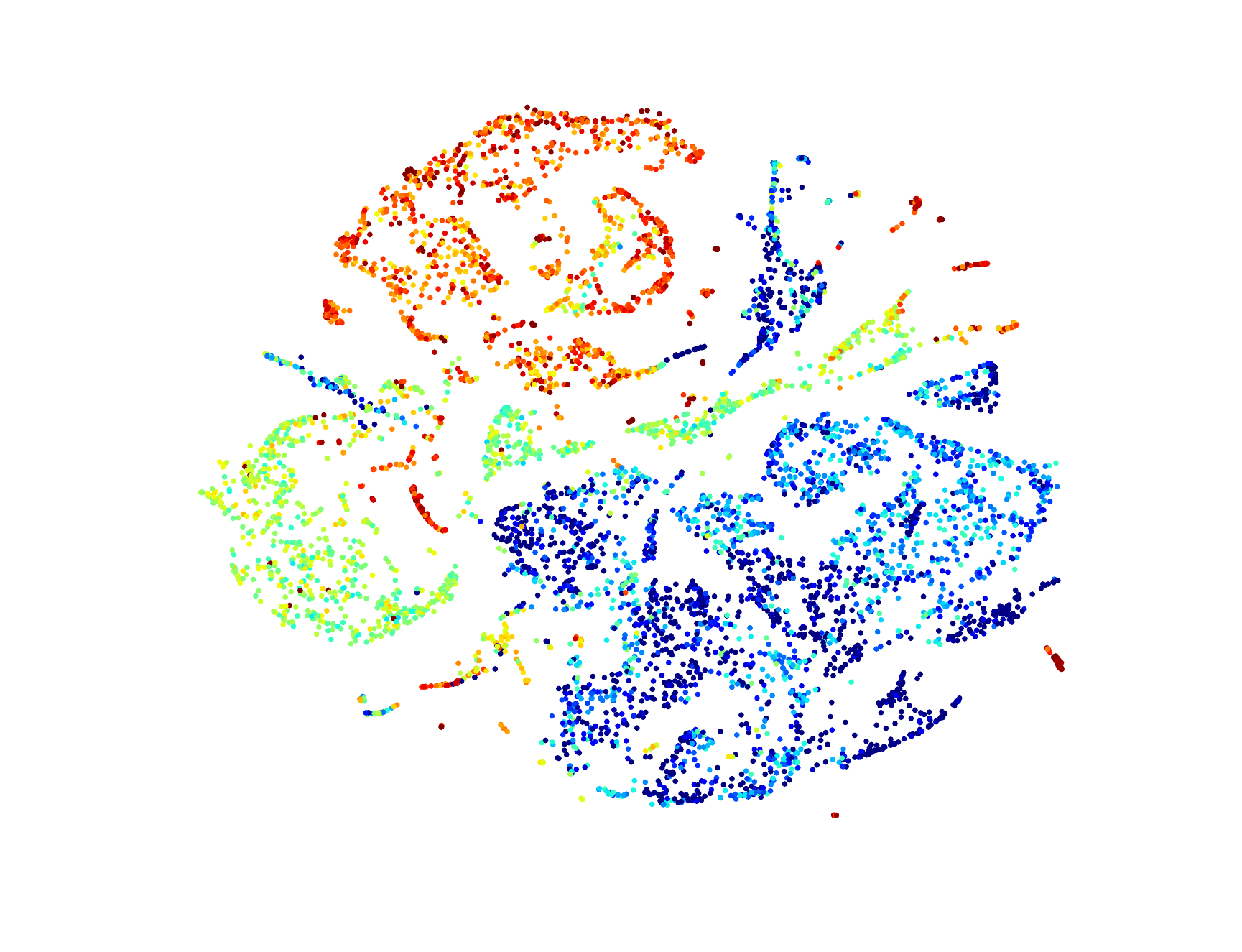}\\
(e) \# communities.
\end{minipage}
\begin{minipage}[t]{0.44\textwidth}
\centering
\includegraphics[width=0.98\textwidth]{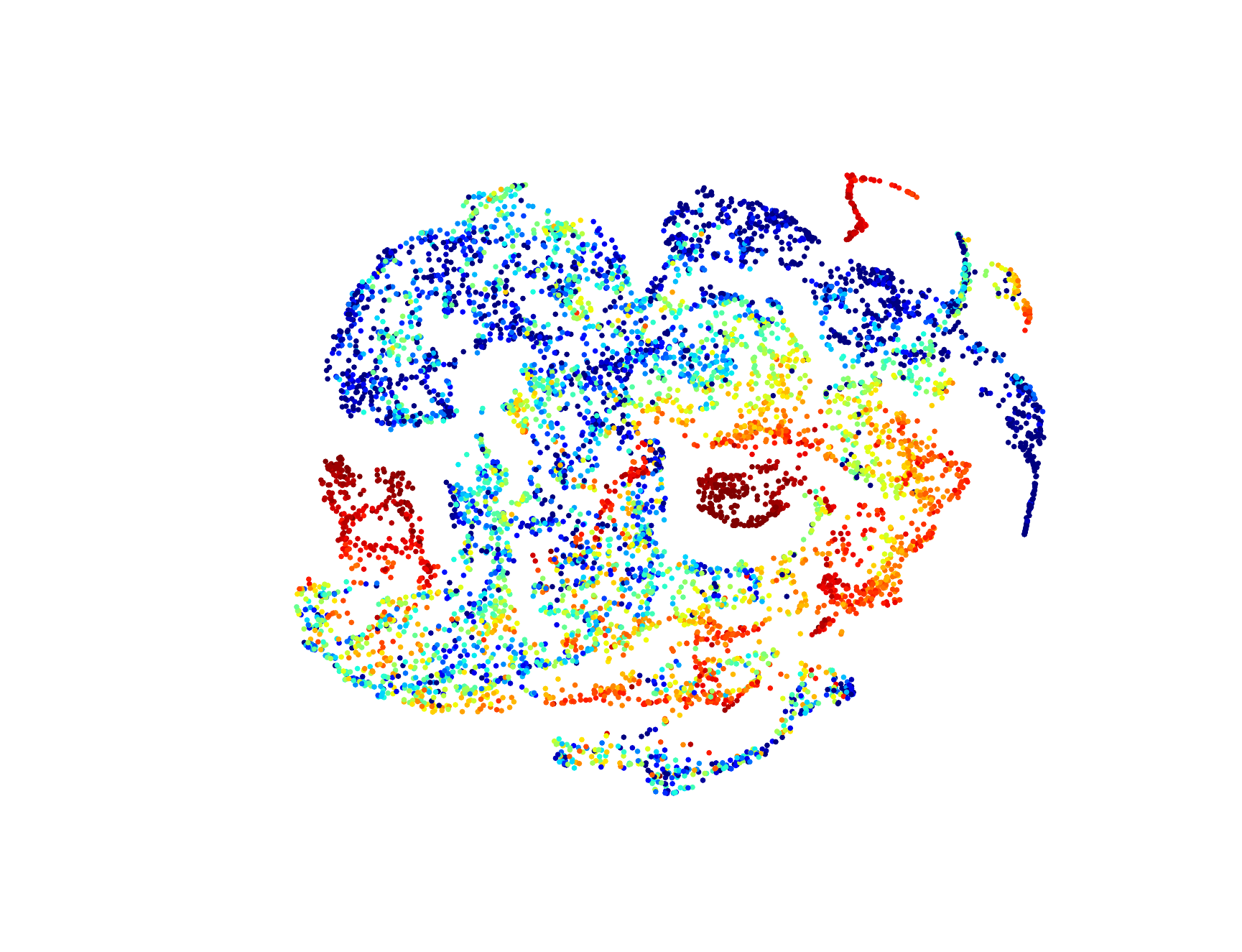}\\
(d) \# communities.
\end{minipage}
\end{minipage}
\begin{minipage}[t]{0.44\textwidth}
\centering
\begin{minipage}[t]{0.44\textwidth}
\centering
\includegraphics[width=0.90\textwidth]{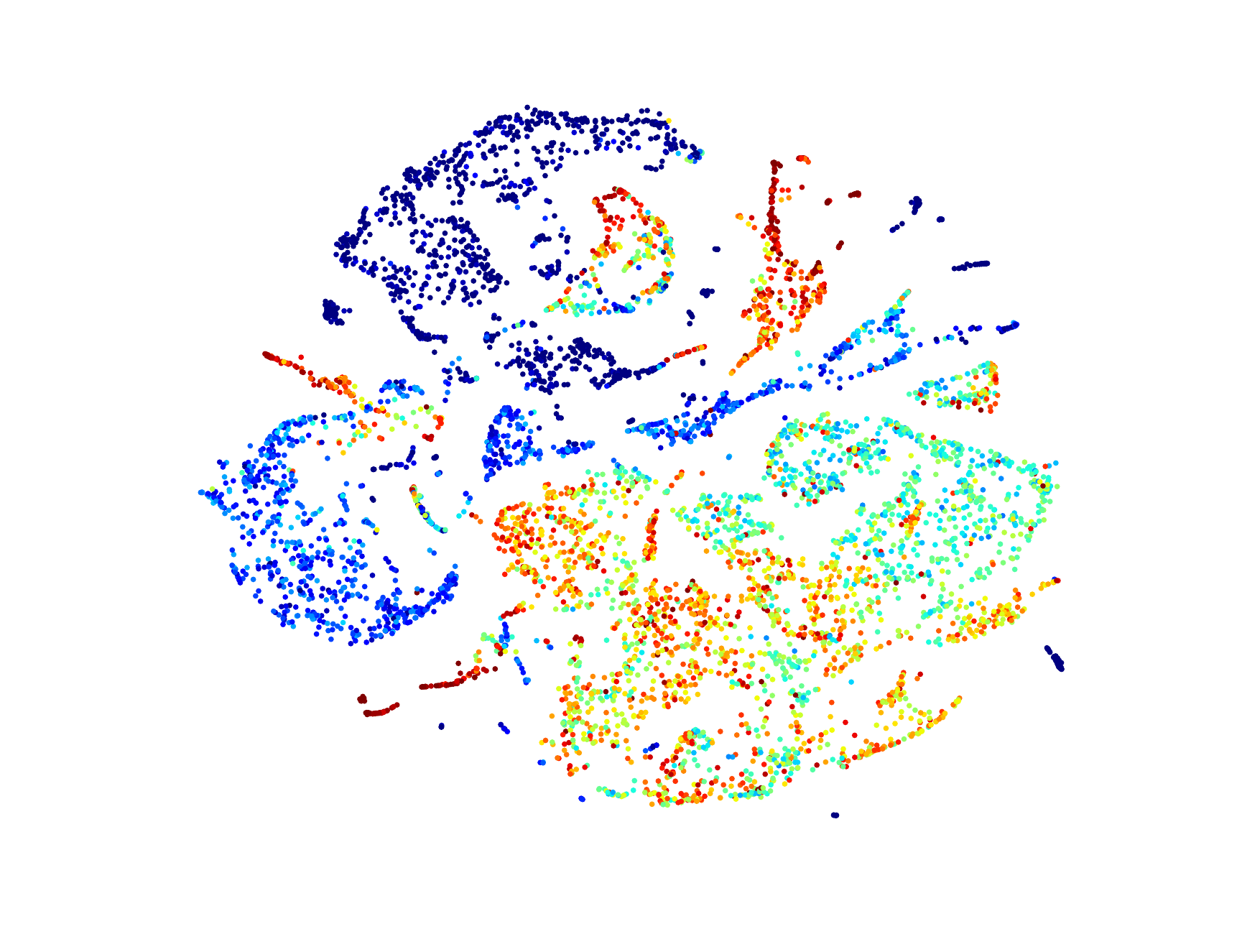}\\
(c) Edge density.
\end{minipage}
\begin{minipage}[t]{0.45\textwidth}
\centering
\includegraphics[width=0.98\textwidth]{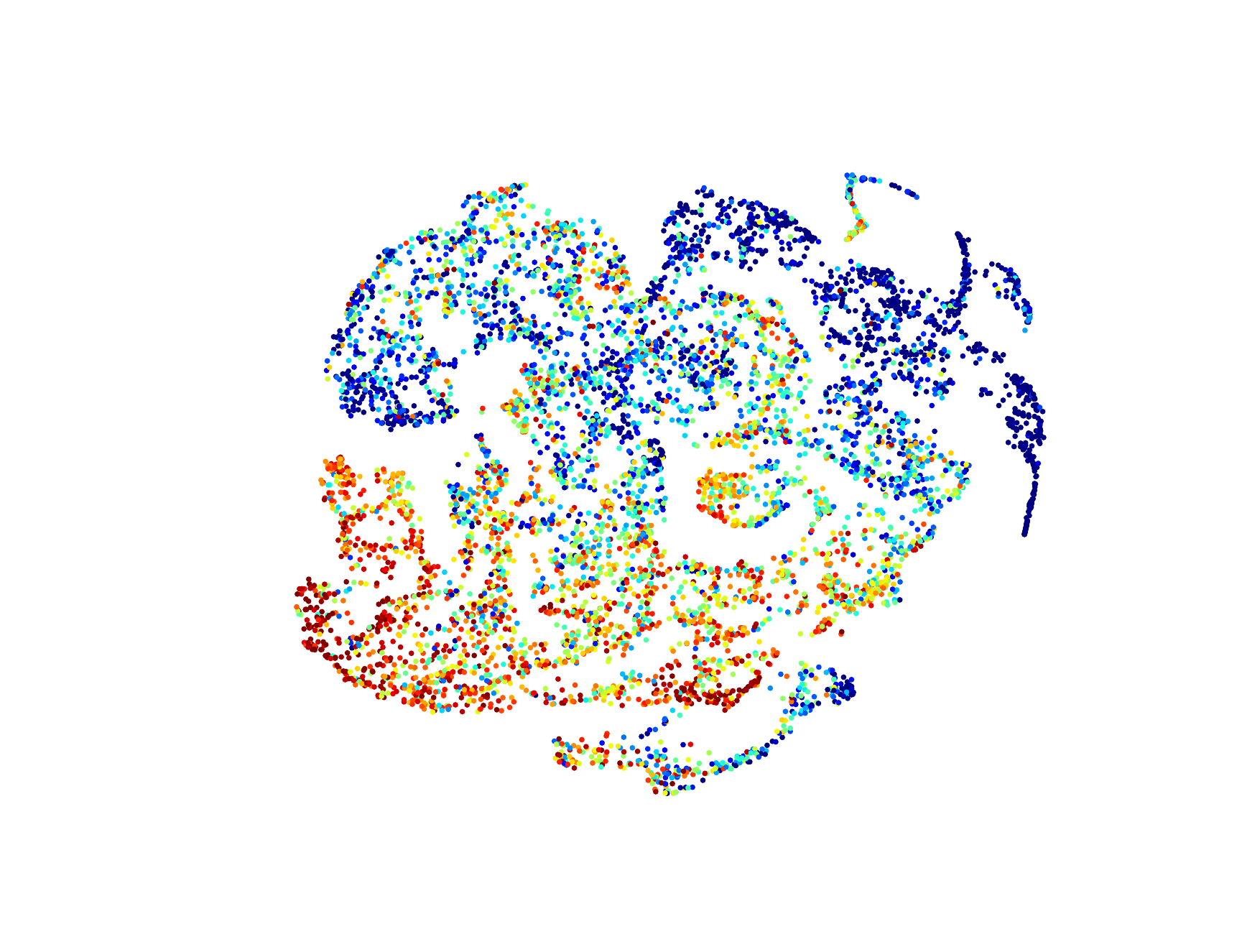}\\
(f) Avg. degree in $\mathcal{G}$.
\end{minipage}
\end{minipage}
\begin{minipage}[t]{0.44\textwidth}
\centering
\begin{minipage}[t]{0.44\textwidth}
\centering
\includegraphics[width=0.90\textwidth]{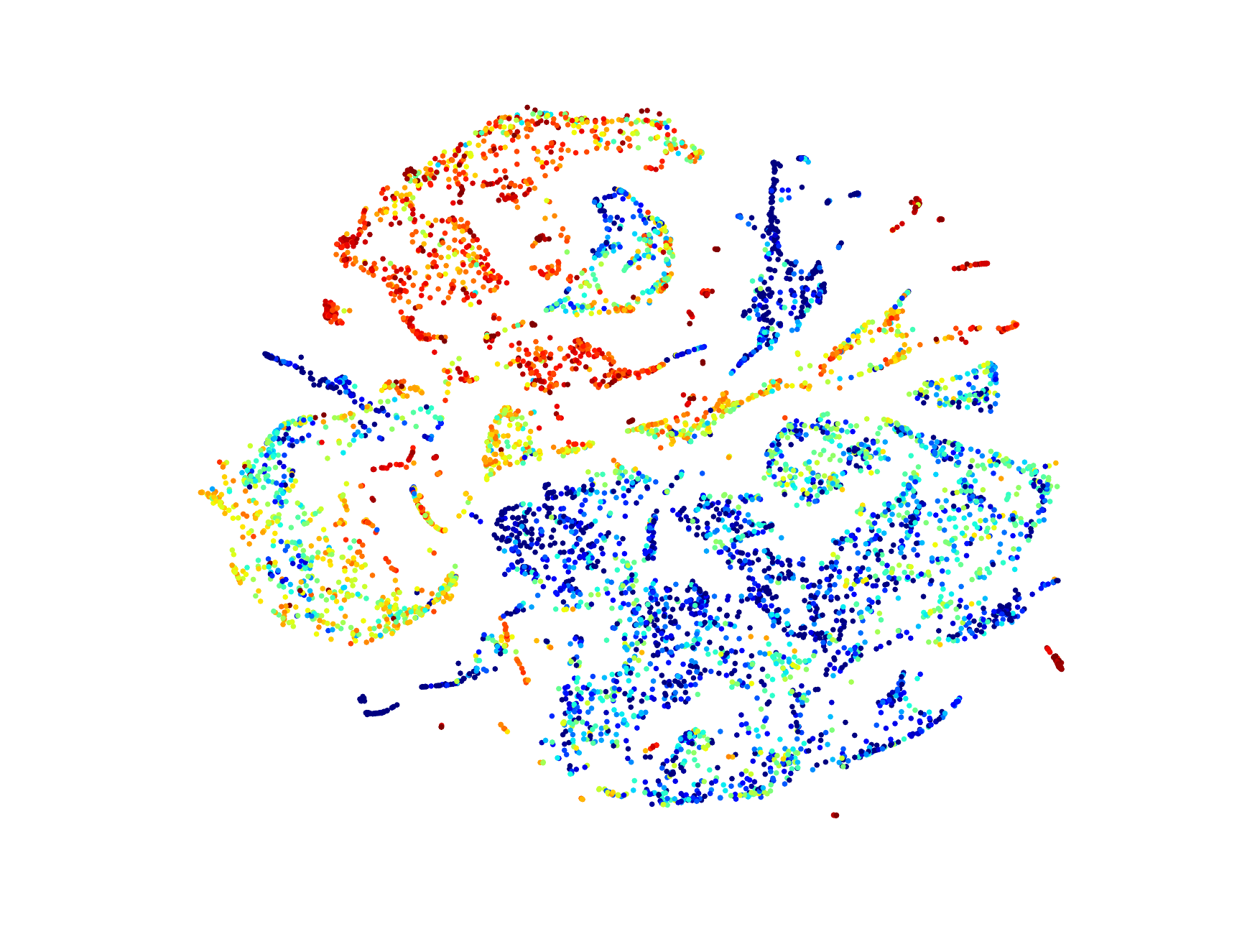}\\
(g) \# leaf nodes.
\end{minipage}
\begin{minipage}[t]{0.45\textwidth}
\centering
\includegraphics[width=0.98\textwidth]{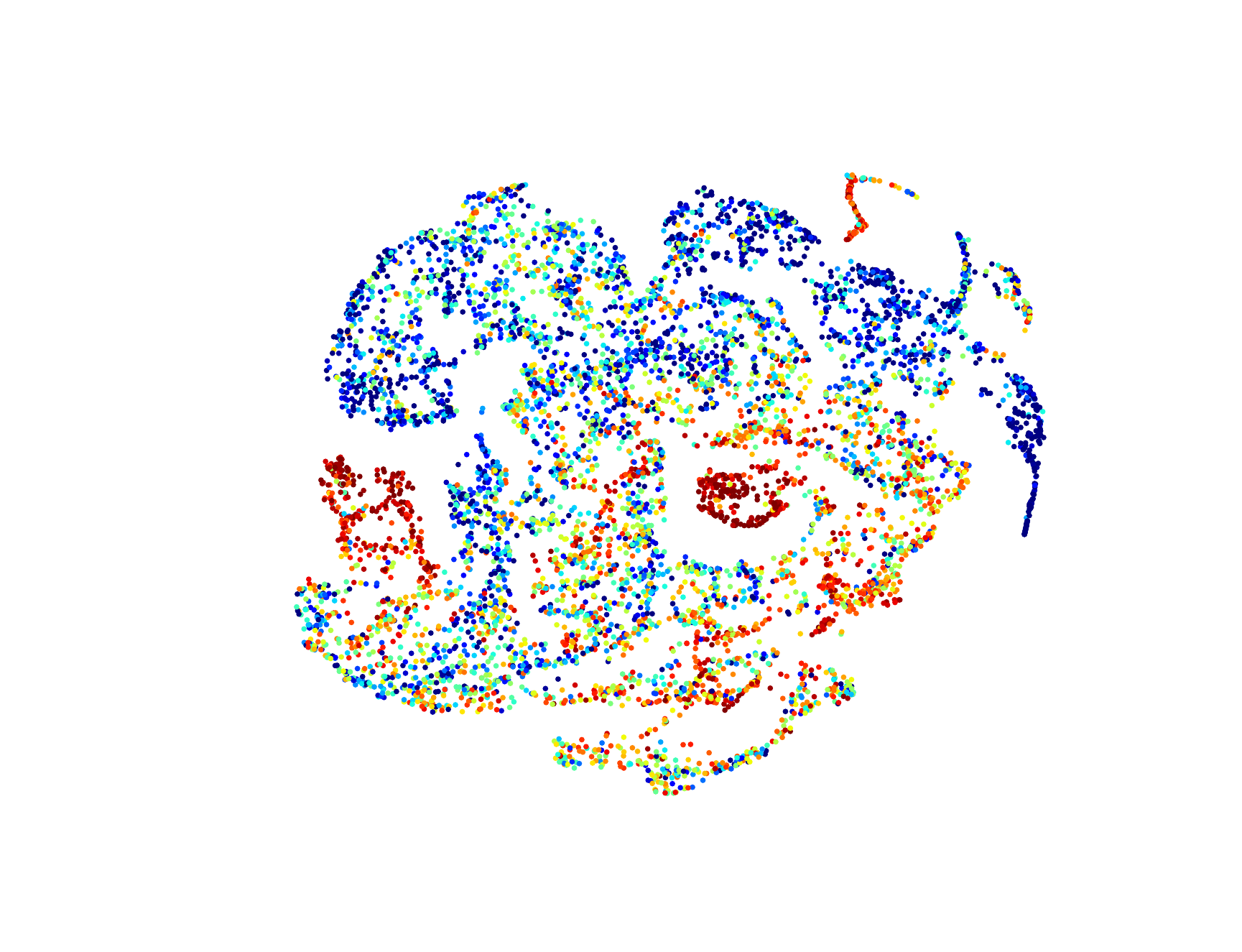}\\
(h) \# edges.
\end{minipage}
\end{minipage}
\begin{minipage}[t]{0.44\textwidth}
\centering
\begin{minipage}[t]{0.44\textwidth}
\centering
\includegraphics[width=0.90\textwidth]{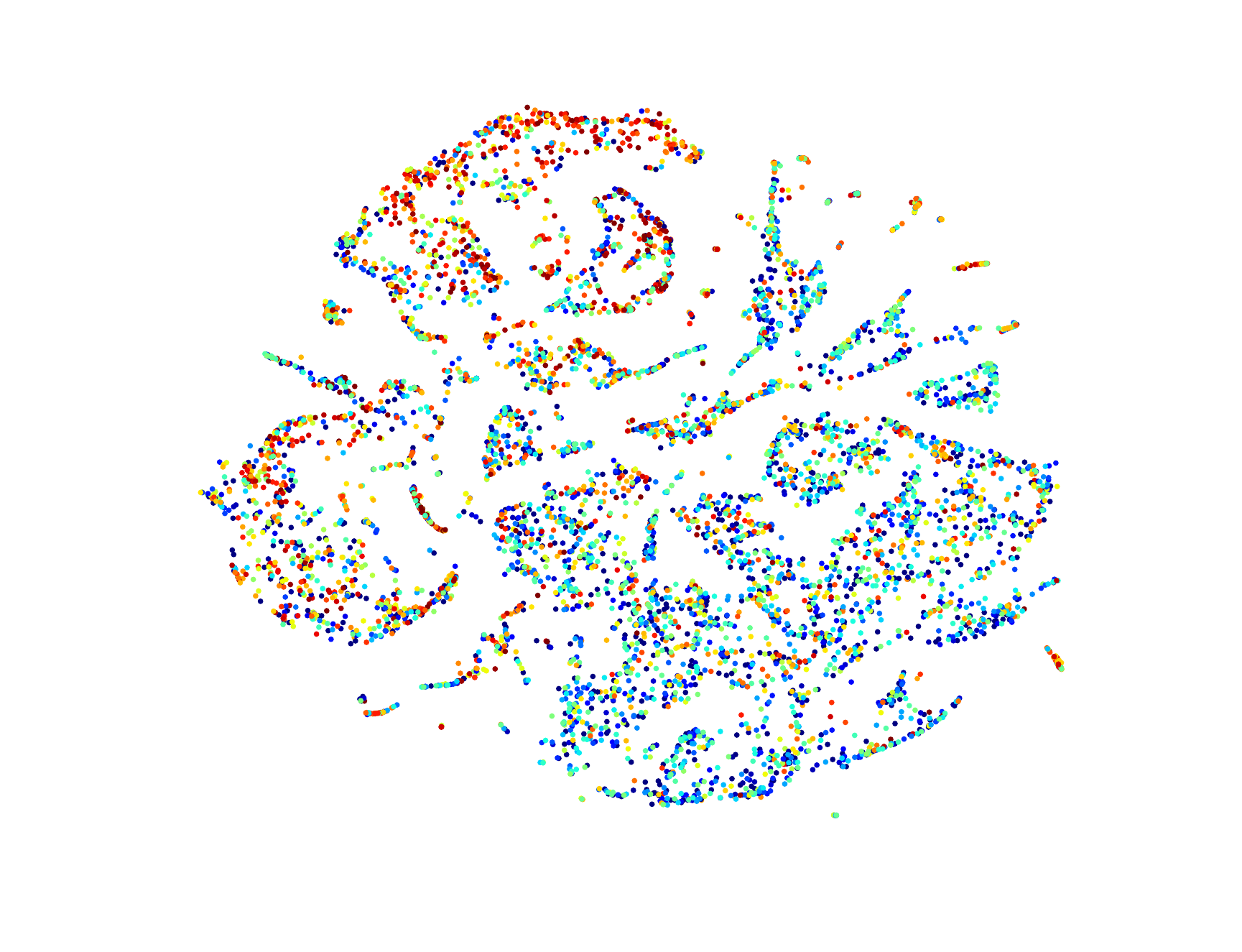}\\
(i) Increment of diffusion size on \textsc{Twitter}.
\end{minipage}
\begin{minipage}[t]{0.44\textwidth}
\centering
\includegraphics[width=0.98\textwidth]{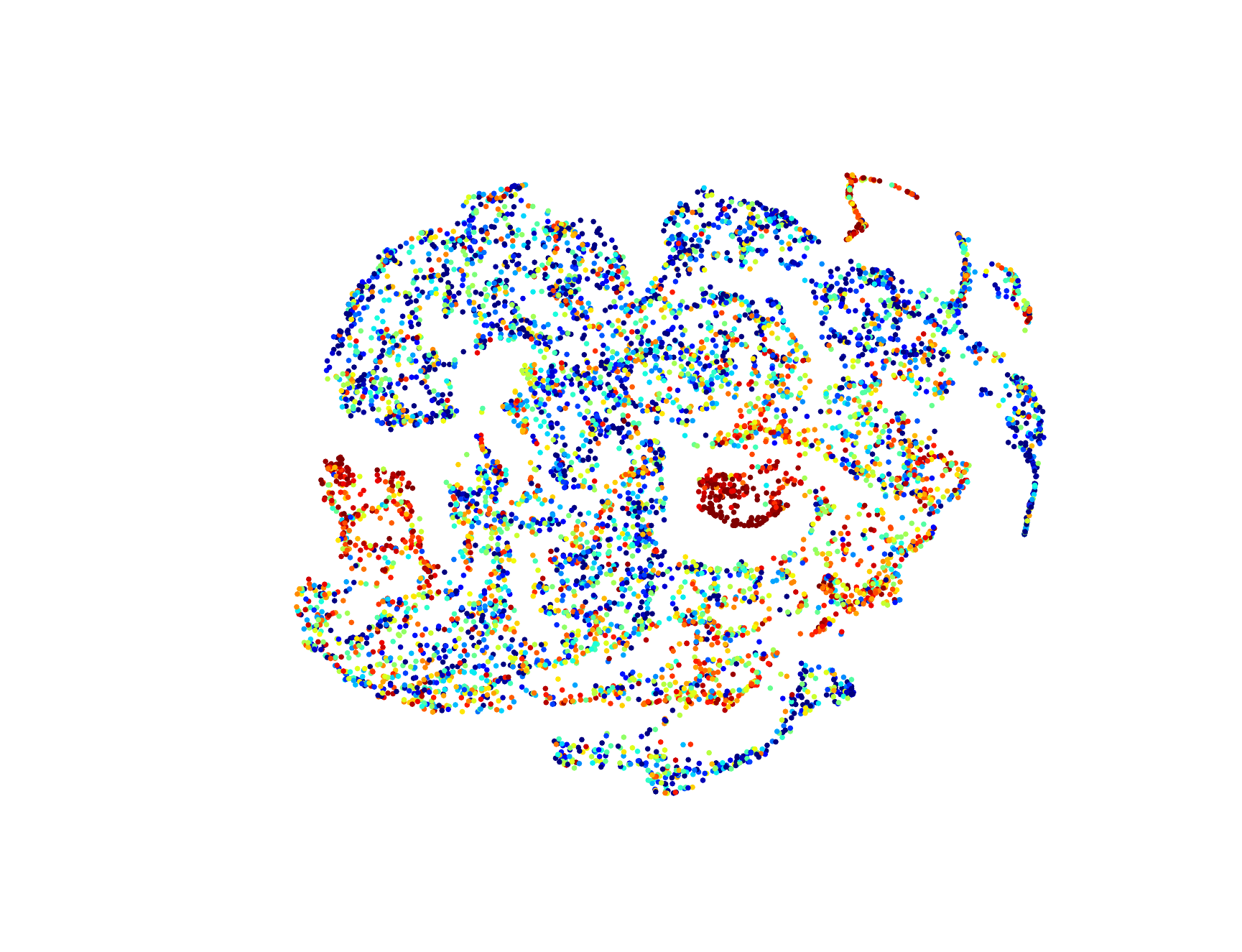}\\
(j) Increment of diffusion size on \textsc{AMiner}.
\end{minipage}
\caption{Feature visualization. Every point is a cascade graph in \textit{test} set. %The layout is produced from high-level representations of DeepCas, 
Every layout is colored (red: high, blue: low) using \textit{hand-crafted} network properties or the ground-truth, labeled under each subfigures. The left column displays graphs from \textsc{Twitter}, while the right column shows \textsc{AMiner}.}
\label{fig:feature_analysis}
\end{minipage}
\end{figure}

To do this, we select a few hand-crafted features which are computed for the feature based baselines. %Note that we work only on test set, as we care more about the prediction performance. 
These features characterize either global or local network properties, and are listed in Figure~\ref{fig:feature_analysis}. 
In each subfigure, we layout the cascade graphs as data points in the test set to a 2-D space by feeding their vector representations output by the last hidden layer of DeepCas to t-SNE~\cite{blondel2008fast}, a commonly used visualization algorithm. 
%The t-SNE algorithm projects feature vectors into a 2-dimensional space, where similar 
Cascade graphs with similar vector representations are placed closely. 
%The visualizations of data sets Tweets and Papers are displayed in Figure~\ref{fig:feature_analysis}. 
To connect the hand-crafted features with the learned representations, we color each cascade graph (a point in the 2-D visualization) by the values of each feature (e.g., network density). If we eyeball a pattern of the distribution of colors in this 2-D layout, it suggests a connection between the learned representation and that network property. We also color the layout by the ground-truth labels (increment of cascade size). If the color distribution of labels looks somewhat correlated with the color distribution of a network property, we know this property attributes to cascade prediction, although not through a hand-crafted feature. 

As we observe, DeepCas could capture structural properties like the number of open, closed triangles, and the number of communities. For example, in the Figure~\ref{fig:feature_analysis} (e), the points (cascade graphs) clustered to the bottom right have the fewest communities, while graphs in the top left have the most. Cascade graph with a larger number of communities implies that many early adopters may lie in between bigger communities, which are likely to be structural holes in the global network. In literature \cite{burt2000network}, nodes spanning structural holes are likely to gain social capital, promoting the growth of its ego-net. Indeed, when we compare the color scheme of \ref{fig:feature_analysis}(g) with \ref{fig:feature_analysis}(i), we can see that the number of communities in a cascade graph is indeed positively correlated with its growth.  

Figure~\ref{fig:feature_analysis} (f) plots the average global degree of nodes in each cascade graph. The pattern suggests that DeepCas not only captures the structural information from individual cascade graphs, but also incorporates the global information into the graph representation. How did this happen? Although we did not explicitly represent the global network $\mathcal{G}$ (or the frontier graphs), DeepCas is likely to learn useful global network information from the many cascade graphs in training (similar to a model that captures collection-level information from the input of many individual documents), and incorporate it into the high-level representation of a cascade graph. 

Some additional observations can be made from Figure~\ref{fig:feature_analysis}. First, as the number of open and closed triangles are actually important features used for graph prediction tasks~\cite{ugander2012structural}, we can see that DeepCas has automatically learned these useful features without human input. Second, since edge density is a function of the number of edges and nodes, DeepCas learns not only the number of edges and nodes (we do not show the node property in Figure~\ref{fig:feature_analysis}, but this is true), but also their none-linear relationship that involves division.

\section{Discussion and Conclusion}

We present the first end-to-end, deep learning based predictor of information cascades. A cascade graph is first represented as a set of random walk paths, which are piped through a carefully designed GRU neural network structure and an attention mechanism to predict the future size of the cascade. The end-to-end predictor, DeepCas, outperforms feature-based machine learning methods and alternative node embedding and graph embedding methods.

While the study adds another evidence to the recent successes of deep learning in a new application, social networks, we do wish to point the readers to a few more interesting implications. First, we find that linearly combined, hand-crafted features perform reasonably well in cascade prediction, which outperform a series of node embedding, graph embedding, and suboptimal deep learning methods. Comparing to other data mining domains, social network is a field where there exists rich theoretical and empirical domain knowledge. Carefully designed features inherited from the literature are already very powerful in capturing the critical properties of networks. The benefit of deep learning in this case really comes from the end-to-end procedure, which is likely to have learned high-level features that just better represent these network properties. Comparing to deep learning methods, feature-based methods do have their advantages (if the right features are identified), as both the results and the importance of features are easier to interpret. For social network researchers, it is perhaps a good idea to interpret DeepCas as a way to test the potential room to improve cascade prediction, instead of as a complete overturn of the existing practice. Indeed, it is intriguing to pursue how to design better measurements of the classical network concepts (e.g., communities and centrality), based on the results of DeepCas.   

Another interesting finding is that different random walk strategies perform better and worth in different scenarios, and all better than bag of node embeddings. This is where prior knowledge in social networks literature may kick in, by incorporating various contagion/diffusion processes to generate initial representations of cascade networks. How to choose from multiple cascading processes itself is an interesting question of reinforcement learning. 

Finally, to make our conclusion clean and generalizable, we only utilized the network structure and node identities in the prediction. It is interesting to incorporate DeepCas with other types of information when they are available, e.g., content and time series, to optimize the prediction accuracy on a particular domain.

\newpage
\bibliographystyle{abbrv}
\bibliography{survey}  % sigproc.bib is the name of the Bibliography in this case

\end{document}